\documentclass[twocolumn,english,prx, eqsecnum]{revtex4-2}
\usepackage[T1]{fontenc}
\usepackage[latin9]{inputenc}
\usepackage{color}
\usepackage{bm}
\usepackage{amsmath}
\usepackage{amssymb}
\usepackage{graphicx}

\makeatletter
\usepackage{babel}

\usepackage{babel}

\makeatother

\usepackage{babel}
\begin{document}
\title{Time evolution with symmetric stochastic action}
\author{Peter~D.~Drummond}
\affiliation{Centre for Quantum and Optical Science, Swinburne University of Technology,
Melbourne 3122, Australia,}
\affiliation{Institute of Theoretical Atomic, Molecular and Optical Physics (ITAMP),
Harvard University, Cambridge, Massachusetts, USA.}
\affiliation{Weizmann Institute of Science, Rehovot, Israel.}
\email{peterddrummond@gmail.com}

\selectlanguage{english}%
\begin{abstract}
The time-evolution of quantum fields is shown to be equivalent to
a time-symmetric Fokker-Planck equation. Results are obtained using
a Q-function representation, including fermion-fermion, boson-boson
and fermion-boson interactions with linear, quadratic, cubic and quartic
Hamiltonians, typical of QED and many other cases. For local boson-boson
coupling, the resulting probability distribution is proved to have
a positive, time-symmetric path integral and action principle, leading
to a forward-backward stochastic process in both time directions.
The solution corresponds to a c-number field equilibrating in an additional
dimension. Paths are stochastic trajectories of fields in space-time,
which are samples of a statistical mechanical steady-state in a higher
dimensional space.  We derive numerical methods and examples of solutions
to the resulting stochastic partial differential equations in a higher
time-dimension, giving agreement with examples of simple bosonic quantum
dynamics. This approach may lead to useful computational techniques
for quantum field theory, as well as to new ontological models of
physical reality.
\end{abstract}
\maketitle

\section{Introduction}

 The role that time plays in quantum mechanics is a deep puzzle in
physics, since quantum measurement appears to preferentially choose
a particular time direction, via the projection postulate. This, combined
with the Copenhagen interpretation that only macroscopic measurements
are real, has led to many quantum paradoxes. Here, we derive a time-symmetric,
stochastic quantum action principle to help resolve these issues,
extending Dirac's idea \citep{dirac1938pam} of future-time boundary
conditions to the quantum domain. In the cases treated here, quantum
field dynamics is shown to be equivalent to a time-symmetric stochastic
process. This can be solved by equilibration in the quasi-time of
a higher dimensional space, with a genuine probability. There are
potentially useful computational consequences, since a quantum action
principle with a real exponent has no phase problem. The resulting
stochastic trajectories also have an ontological interpretation \citep{drummond2020retrocausal}.

The theory uses the Q-function of quantum mechanics \citep{Husimi1940,Hillery_Review_1984_DistributionFunctions,drummond2014quantum},
which is the expectation value of a coherent state projector. It is
a well-defined and positive distribution for any bosonic or fermionic
quantum density matrix \citep{FermiQ}. In Hamiltonians with up to
quartic interactions, the dynamical equation has a time-symmetric
Fokker-Planck form. In bosonic cases this is shown to have a zero
trace diffusion matrix. There is a quantum action principle for diffusion
in positive and negative time-directions simultaneously, equivalent
to a time-symmetric stochastic process. The result is time-reversible
and\emph{ }non-dissipative, explaining how quantum evolution can be
inherently random yet time-symmetric.

Using stochastic bridge theory \citep{schrodinger1931uber,hairer2007analysis,drummond2017forward},
the Q-function time-evolution is proved to correspond to the steady-state
of a diffusion equation in an extra dimension. Thus, stochastic equilibration
of a c-number field in five dimensions gives rise to quantum dynamics
in four dimensional space-time. This shows that c-number fields in
higher dimensions can behave quantum-mechanically. No imaginary-time
propagation is required, and the description is probabilistic. The
relationship to quantum ontology and measurement theory is given elsewhere
\citep{reid2017interpreting,drummond2020retrocausal}.

The stochastic fields used here propagate retrocausally as well as
in a positive time-direction. Time symmetric evolution was originally
proposed by Tetrode in classical electrodynamics \citep{tetrode1922causal}.
Dirac used the approach to obtain an elegant theory of classical radiation
reaction \citep{dirac1938pam}, which was extended by Feynman and
Wheeler \citep{wheeler1945interaction}. Time-reversible methods have
been studied in quantum physics \citep{aharonov1964time,cramer1980generalized,pegg1982time,pegg2002quantum},
the philosophy of science \citep{price2008toy}, and used to explain
Bell violations \citep{argaman2010bell}. Here, we use this general
approach to analyze \emph{interacting} fields, thus giving time-symmetric
quantum physics a strong foundation.

By comparison, the Fenyes-Nelson approach to stochastic quantum evolution
\citep{fenyes1952wahrscheinlichkeitstheoretische,nelson1966derivation},
does not have a constructive interpretation \citep{grabert1979quantum}.
The approach of stochastic quantization \citep{parisi1981perturbation}
uses imaginary time. Such methods have the drawback that analytic
continuation to real time dynamics can be intractable \citep{silver1990maximum,feldbrugge2017lorentzian}.
The mathematical technique used here combines the Wiener-Stratonovich
stochastic path integral \citep{wiener1930generalized,stratonovich1971probability},
with Schr\"odinger's \citep{schrodinger1931uber} idea of a stochastic
bridge in statistical mechanics, as generalized by later workers.
The resulting equilibrium is equivalent to quantum dynamics.

All quantum effects are retained in this approach \citep{drummond2019q}.
This is not unexpected, because quantum absorber theory, with similar
time-reversed propagation, even has Bell violations \citep{pegg1982time}.
There is a related approach of reciprocal diffusion due to Bernstein
\citep{bernstein1932liaisons} and Zambrini \citep{zambrini1986stochastic}.
The focus of this paper is to understand quantum dynamics and measurement
using stochastic methods. This is important to fundamental applications
to quantum measurement theory \citep{drummond2020retrocausal}. In
addition, stochastic methods scale well for large systems, which may
help to compute exponentially complex many-body dynamics.

Historically, the Kaluza-Klein theory of electromagnetism \citep{kaluza1921unitatsproblem,klein1926quantentheorie,overduin1997kaluza}
first introduced a fifth dimension. String theory \citep{horowitz2005spacetime,mohaupt2003introduction},
as well as the Randall-Sundrum \citep{randall1999alternative} and
Gogberashvili \citep{Gogberashvili2000} approach to the hierarchy
problem all use extra dimensions. As with these approaches, one can
usefully visualize the extra dimensions as real extensions of space-time.
In the theory presented here, the extra dimension is time-like and
non-compact. Just as in `flatland' \citep{abbott2006flatland}, the
location of observers defines the extra coordinate. However, our results
only make use of an extra dimension as a computational tool, so this
is not essential to the physical interpretation.

Quantum dynamical problems arise in many fields, from many-body theory
to cosmology. The utility of the path integral derived here is that
it is real, not imaginary \citep{feynman2010quantum}. Other methods
exist for quantum dynamics. These include mean field theory, perturbation
theory, variational approaches \citep{cosme2016center}, standard
phase-space methods \citep{Hillery_Review_1984_DistributionFunctions}
and the density matrix renormalization group \citep{White:1992}.
Each has its own drawbacks, however. The time-symmetric techniques
given here use a different approach, as well as providing a model
for a quantum ontology.

To demonstrate these results, a general quartic quantum field Hamiltonian
is introduced. This corresponds to common quantum models, including
QED, Bose-Hubbard, Fermi-Hubbard, Ising, quantum circuit and parametric
Hamiltonians. The corresponding Q-function dynamics satisfies a time-symmetric
Fokker-Planck equation in all cases. Further extensions of the Q-function
method would be required to treat the bosonic sector in QCD. Such
Hamiltonians are closely related to the classes treated here. The
common feature is that they have a cubic or quartic interaction term.{{}
In the boson-boson and boson-fermion cases it is shown that there
is a zero trace diffusion. A detailed treatment is given of the simplest
boson-boson case. This leads to a time-symmetric stochastic differential
equations, a quantum action principle and a probabilistic path integral.
A solution is obtainable through positive diffusion in a higher dimension.
Elementary examples and numerical solutions are obtained. Results
are compared with exactly soluble cases.}

{The content of this paper is as follows. Section
\ref{sec:Q-functions} derives properties of Q-functions and time-symmetric
Fokker-Planck equations, proving that there is a traceless diffusion
matrix for boson-boson and fermion-boson quantum field theories. Section
\ref{sec:Bidirectional-stochastic-bridges} defines quantum trajectories
and an action principle using time-symmetric stochastic equations.
Section \ref{sec:Q-function-and-path-integral} shows that the resulting
real path integral is equivalent to a Q-function time-evolution. Section
\ref{sec:Extra-dimensions} treats extra dimensions, and shows how
the classical limit is regained. Section \ref{sec:Quadratic-Hamiltonian-Examples}
gives examples and numerical results. Finally, section \ref{sec:Summary}
summarizes the paper.}

\section{Q-functions\label{sec:Q-functions}}

Phase-space representations in quantum mechanics allow efficient treatment
of quantum systems via probabilistic sampling \citep{drummond2016quantum}.
These general methods are related to coherent states \citep{Glauber_1963_P-Rep}
and Lie groups \citep{Perelomov_1972_Coherent_states_LieG}, which
introduce a continuous set of parameters in Hilbert space. Results
for bosonic and fermionic fields are summarized below. Q-functions
\citep{Husimi1940} can also be used for spins \citep{Arecchi_SUN,Drummond1984PhysLetts}
but this is essentially a subset of the results given here.

The Q-function is probabilistic and defined in real time. Yet it does
$not$ have a traditional stochastic interpretation, since unitary
evolution can generate diffusion terms that are not positive-definite
\citep{zambrini2003non}. An earlier method of treating this was to
double the phase-space dimension to give a positive diffusion \citep{Drummond_Gardiner_PositivePRep}.
This is usually applied to normal ordering \citep{Glauber_1963_P-Rep},
but the corresponding distribution is non-unique, and is most useful
for damped systems \citep{Drummond:1999_QDBEC} or short times \citep{Carter:1987,Deuar:2007_BECCollisions,Corney_2008}.
With undamped systems however, doubling phase-space gives sampling
errors that increase with time \citep{Deuar2006a,Deuar2006b}. Rather
than using this earlier approach, a positive diffusion is obtained
here through equilibration in an extra space-time dimension.

\subsection{General definition of a Q-function}

The general abstract definition of a Q-function \citep{FermiQ} is:
\begin{equation}
Q(\bm{\lambda},t)=Tr\left\{ \hat{\Lambda}\left(\bm{\lambda}\right)\hat{\rho}\left(t\right)\right\} \,,\label{eq:Q-definition}
\end{equation}
where $\hat{\rho}\left(t\right)$ is the quantum density matrix, $\hat{\Lambda}\left(\bm{\lambda}\right)$
is a positive-definite operator basis, and $\bm{\lambda}$ is a point
in phase-space with an integration measure $d\bm{\lambda}$ such that
the identity operator $\hat{I}$ can be expanded as: 
\begin{equation}
\hat{I}=\int\hat{\Lambda}\left(\bm{\lambda}\right)d\bm{\lambda}\,.\label{eq:completeness}
\end{equation}

Provided $\hat{\rho}\left(t\right)$ is normalized in the standard
way so that $Tr\left[\hat{\rho}\left(t\right)\right]=1$, the Q-function
is positive and normalized to unity:
\begin{equation}
\int d\bm{\lambda}Q\left(\bm{\lambda}\right)=1\,.\label{Q-normalization}
\end{equation}
 This satisfies the requirements of a probability. The basis function
$\hat{\Lambda}\left(\bm{\lambda}\right)$ does not project the eigenstates
of a Hermitian operator, and therefore the quantum dynamical equations
differ from those for orthogonal eigenstates. 

Quantum fields are defined with an $n_{d}$-dimensional space-time
coordinate $r$, where $r=\left(r^{1},\ldots r^{n_{d}}\right)=\left(\bm{r},t\right)$.
The fields can be expanded using $M_{b}$ bosonic annihilation and
creation operators $\hat{a}_{i},\hat{a}_{i}^{\dagger}$ and $M_{f}$
fermionic operators $\hat{b}_{i},\hat{b}_{i}^{\dagger}$. The internal
indices $i$ may include $N_{b}$ ($N_{f}$) internal degrees of freedom
like spin or charm, for $M_{b}/N_{b}$ and $M_{f}/N_{f}$ spatial
modes respectively, each of which describe excitations on a lattice
or single-particle eigenmodes.

As a result, the corresponding phase-space coordinate $\bm{\lambda}=\left(\bm{\alpha},\underline{\underline{\xi}}\right)$
has both fermionic and bosonic amplitudes, and the basis function
$\hat{\Lambda}\left(\bm{\lambda}\right)$ is a product of fermionic
and bosonic operators, so that:
\begin{equation}
\hat{\Lambda}\left(\bm{\lambda}\right)=\prod_{b,f}\hat{\Lambda}_{b}\left(\bm{\alpha}\right)\hat{\Lambda}_{f}\left(\underline{\underline{\xi}}\right).
\end{equation}
We now briefly recall the properties of the basis operators, $\hat{\Lambda}_{b}\left(\bm{\alpha}\right)$
and $\hat{\Lambda}_{f}\left(\underline{\underline{\xi}}\right)$.
For simplicity we assume one bosonic and/or one fermionic field here,
noting that there may be more than one species labeled $b,f$ if there
are conservation laws. They must satisfy the completeness condition,
Eq (\ref{eq:completeness}). The basis is not orthogonal, and it is
generally essential to employ non-orthogonal bases and Lie group theory
in order to obtain differential and integral identities.

For bosonic fields, $\hat{\Lambda}_{b}$ is proportional to a coherent
state projector \citep{Glauber_1963_P-Rep},
\begin{equation}
\hat{\Lambda}_{b}\left(\bm{\alpha}\right)\equiv\left|\bm{\alpha}\right\rangle _{c}\left\langle \bm{\alpha}\right|_{c}/\pi^{M_{b}}.
\end{equation}
The state $|\bm{\alpha}\rangle_{c}$ is a normalized Bargmann-Glauber
\citep{Bargmann:1961,Glauber_1963_P-Rep} coherent state with $\hat{a}_{i}|\bm{\alpha}\rangle_{c}=\alpha_{i}|\bm{\alpha}\rangle_{c}$
and $\hat{\bm{\psi}}\left(\bm{r}\right)|\bm{\alpha}\rangle_{c}=\bm{\psi}\left(\bm{r}\right)|\bm{\alpha}\rangle_{c}$,
where $\bm{\alpha}$ is an $M_{b}$-dimensional complex vector of
coherent field mode amplitudes and $\bm{\psi}\left(\bm{r}\right)$
is the corresponding coherent field. Here, Latin indices $i,j,k,\ell$
are summed up to $M_{b}$, with bold vectors and matrices. On Fourier
transforming to position space, $Q\left[\bm{\psi}\right]$ in field
space is a functional of the complex field amplitudes $\bm{\psi}\left(\bm{r}\right)$
\citep{Steel1998,opanchuk_2013}. 

For fermionic fields, the Gaussian operator $\hat{\Lambda}_{f}$ is
given in the Majorana representation by~\citep{Riashock2018,joseph2018phase}
\begin{equation}
\hat{\Lambda}_{f}\left(\underline{\underline{\xi}}\right)=\mathcal{N}\left(\underline{\underline{\xi}}\right):\exp\Biggl[-i\hat{\underline{\gamma}}^{T}\left[\underline{\underline{i}}-\underline{\underline{i}}\left(\underline{\underline{i}}\underline{\underline{\xi}}+\underline{\underline{I}}\right)^{-1}\right]\hat{\underline{\gamma}}/2\Biggr].\label{eq:MajOpX}
\end{equation}
The fermionic $\underline{\underline{\xi}}$ coordinates are real
antisymmetric $2M_{f}\times2M_{f}$ matrices with an integration measure
$d\underline{\underline{\xi}}=\prod_{m<n}d\xi_{mn}$, and an integration
domain such that $\left(\underline{\underline{\xi}}^{2}+\underline{\underline{I}}\right)$
is a positive semi-definite matrix. Latin indices $m,n,o,p$ for extended
vectors are summed up to $2M_{f}$, with underlined vectors and matrices.
The normalization factor is $\mathcal{N}\left(\underline{\underline{\xi}}\right)=2^{-M_{f}}\sqrt{\det\left[\underline{\underline{i}}-\underline{\underline{\xi}}\right]}$,
where
\begin{equation}
\underline{\underline{i}}=\left[\begin{array}{cc}
\mathbf{0} & \mathbf{\mathbf{I}}\\
-\mathbf{I} & \mathbf{0}
\end{array}\right].
\end{equation}

Here $\underline{\underline{I}}$ is a $2M_{f}\times2M_{f}$ identity
matrix, $\mathbf{\mathbf{I}}$ is an $M_{f}\times M_{f}$ identity
matrix, and $\hat{\underline{\gamma}}$ is a vector of Majorana operators
with commutators
\begin{equation}
\left\{ \hat{\gamma}_{m},\hat{\gamma}_{n}\right\} =2\delta_{mn}.
\end{equation}
The Majorana operators are obtained using a matrix transformation~\citep{Balian_Brezin_Transformations}
$\underline{\underline{U}}=\left[\begin{array}{cc}
\mathbf{I} & \mathbf{I}\\
-i\mathbf{I} & i\mathbf{I}
\end{array}\right]$, acting on a vector of extended fermionic creation and annihilation
operators, $\hat{\underline{b}}=\left(\hat{\bm{b}}^{T},\hat{\bm{b}}^{\dagger}\right)^{T}$,
so that $\hat{\underline{\gamma}}=\underline{\underline{U}}\hat{\underline{b}}.$
There are $M_{f}\left(2M_{f}-1\right)$ independent fermionic phase-space
variables, owing to antisymmetry.

\subsection{Observables}

Quantum expectations $\left\langle \hat{O}\right\rangle $ of ordered
observables $\hat{O}$ are identical to  probabilistic Q-function
averages $\left\langle O\right\rangle _{Q}$, including corrections
for operator re-ordering if necessary:
\begin{equation}
\left\langle \hat{O}\right\rangle =\left\langle O\right\rangle _{Q}\equiv\int d\bm{\lambda}Q\left(\bm{\lambda}\right)O(\bm{\lambda})\,.
\end{equation}

Here, $\left\langle \right\rangle $ indicates a quantum expectation
value, $\left\langle \right\rangle _{Q}$ is a Q-function phase-space
probabilistic average, and time-arguments are implicit. 

In the bosonic case, the expectation of any observable $\hat{O}$
is obtained by first expanding $\hat{\rho}$ in a generalized P-representation,
$P\left(\alpha,\beta\right)$. This always exists \citep{Drummond1980},
so that for any quantum density matrix $\hat{\rho}$,
\begin{equation}
\hat{\rho}=\int P\left(\bm{\alpha},\bm{\beta}\right)\hat{\Lambda}_{p}\left(\bm{\alpha},\bm{\beta}\right)d\bm{\alpha}d\bm{\beta}\,,\label{eq:Positive-P-expansion}
\end{equation}
where $\hat{\Lambda}_{p}\left(\bm{\alpha},\bm{\beta}\right)$ is an
off-diagonal coherent projector,
\begin{equation}
\hat{\Lambda}_{p}\left(\bm{\alpha},\bm{\beta}\right)=\frac{\left|\bm{\alpha}\right\rangle _{c}\left\langle \bm{\beta}\right|_{c}}{\left\langle \bm{\beta}\right.\left|\bm{\alpha}\right\rangle _{c}}\,.
\end{equation}
We use $d\bm{\alpha}$, $d\bm{\beta}$ to denote $M$ dimensional
complex integration measures, so that if $\bm{\alpha}=\bm{q}+i\bm{p}$,
then $d\bm{\alpha}=d\bm{q}d\bm{p}\equiv d^{M}\bm{q}d^{M}\bm{p}$.
The existence proof \citep{Drummond1980} shows that there is a canonical
probability distribution $P\left(\bm{\alpha},\bm{\beta}\right)$ which
is obtained from the Q-function:
\begin{equation}
P\left(\bm{\alpha},\bm{\beta}\right)=\left(\frac{1}{4\pi}\right)^{M}\exp\left[-\frac{\left|\bm{\alpha}-\bm{\beta}\right|^{2}}{4}\right]Q\left(\frac{\bm{\alpha}+\bm{\beta}}{2}\right)\,.\label{eq:canonical-expansion}
\end{equation}

This can be used to invert the Q-function representation mapping,
to obtain the general operator correspondence function for $\hat{O}$
in the form of an integration over $\bm{\alpha}$:
\begin{equation}
\left\langle \hat{O}\right\rangle \equiv\text{\ensuremath{\int}d\ensuremath{\bm{\alpha}Q^{\alpha}\left(\bm{\alpha}\right)}O\ensuremath{\left(\bm{\alpha}\right)}}=\left\langle O\right\rangle _{Q}.
\end{equation}
To prove this, we use the expansion of $\hat{\rho}$ in Eq (\ref{eq:Positive-P-expansion}),
which gives that:
\begin{equation}
\left\langle \hat{\sigma}\right\rangle _{Q}\equiv\int P\left(\bm{\beta},\bm{\gamma}\right)Tr\left[\hat{O}\hat{\Lambda}_{p}\left(\bm{\beta},\bm{\gamma}\right)\right]d\bm{\beta}d\bm{\gamma}\,.
\end{equation}

Expanding this using the canonical expansion, Eq (\ref{eq:canonical-expansion}),
the c-number function corresponding to $\hat{O}$ is therefore $O\left(\bm{\alpha}\right)$,
where on defining $\bm{\alpha}=\left(\bm{\beta}+\bm{\gamma}\right)/2$,
$\bm{\Delta}=\left(\bm{\beta}-\bm{\gamma}\right)/2$:
\begin{equation}
O\left(\bm{\alpha}\right)=\frac{1}{\pi^{M}}\int e^{-\left|\Delta\right|^{2}}Tr\left[\hat{O}\hat{\Lambda}_{p}\left(\bm{\alpha}+\bm{\Delta},\bm{\alpha}-\bm{\Delta}\right)\right]d\bm{\Delta}\,.
\end{equation}
As an example, particle numbers in the bosonic case are given by introducing
the equivalent c-number function $n\left(\alpha\right)\equiv\left|\alpha\right|^{2}-1$,
so that the quantum and probabilistic averages agree:
\begin{equation}
\left\langle \hat{n}\right\rangle =\left\langle n\left(\alpha\right)\right\rangle _{Q}=\left\langle \left|\alpha\right|^{2}-1\right\rangle _{Q}.\label{eq:number_variable}
\end{equation}

This is a special case of the general identity given above. As another
example, a $p-th$ order anti-normally ordered moment is
\begin{align}
\left\langle \hat{a}_{i1}\ldots\hat{a}_{i_{p}}^{\dagger}\right\rangle  & =\left\langle \alpha_{i_{1}}\ldots\alpha_{i_{p}}^{*}\right\rangle _{Q}.
\end{align}
These operator moments can be of any order $p$.

Similar techniques are available for fermions \citep{FermiQ} and
spins \citep{Arecchi_SUN,Drummond1984PhysLetts}, so this approach
is not restricted to bosonic fields. As emphasized in Dirac's review
paper \citep{Dirac_RevModPhys_1945}, one can calculate any observable
average from a  quasi-distribution, provided the observable is expressed
in terms of a suitable operator ordering. In the bosonic case, it
is the anti-normal ordering of ladder operators that is utilized for
Q-functions.

\subsection{Identities and exact results}

There are several mathematical properties that make this expansion
a useful approach. We first introduce a shorthand notation for differential
operators, $\partial_{i}\equiv\partial/\partial\alpha_{i}$. For convenience
in treating products of less than four boson operators a convention
of defining $\hat{a}_{0}=1=\alpha_{0}$ and $\partial_{0}\equiv0$
is used. 

The following operator correspondences for bosons hold for the unit
terms with zero index as well as for operators:
\begin{eqnarray}
\hat{a}_{j}\hat{\Lambda} & = & \alpha_{j}\hat{\Lambda}\nonumber \\
\hat{\Lambda}\hat{a}_{j}^{\dagger} & = & \alpha_{j}^{*}\hat{\Lambda}\nonumber \\
\hat{a}_{j}^{\dagger}\hat{\Lambda} & = & \left(\partial_{j}+\alpha_{j}^{*}\right)\hat{\Lambda}\nonumber \\
\hat{\Lambda}\hat{a}_{j} & = & \left(\partial_{j}^{*}+\alpha_{j}\right)\hat{\Lambda}.
\end{eqnarray}
For $j=0$, the first two identities are trivial since $\hat{a}_{0}=\alpha_{0}$.
The last two follow immediately, since because $\partial_{0}\equiv0$
, these are simply conjugates of the first two identities. 

To illustrate that this introduces no contradictions, note that for
operator products, the differential identities give the following
result for a commutator:
\begin{align}
\left[\hat{a}_{i},\hat{a}_{j}^{\dagger}\right]\hat{\Lambda} & =\left[\partial_{j},\alpha_{i}\right]\hat{\Lambda}\nonumber \\
\hat{\Lambda}\left[\hat{a}_{i},\hat{a}_{j}^{\dagger}\right] & =\left[\partial_{i}^{*},\alpha_{j}^{*}\right]\hat{\Lambda}.
\end{align}

These identities give the result that $\left[\hat{a}_{i},\hat{a}_{j}^{\dagger}\right]\hat{\Lambda}=\delta_{ij}\hat{\Lambda}$
if $i,j>0$, as required for an operator. If one of $i,j$ are zero,
then the commutator identity vanishes, as it must for a c-number.
The resulting identities have the same form as for the usual annihilation
and creation operators. This allows linear, quadratic and cubic terms
to be treated in a uniform formalism. Einstein summations over $i,j\ldots=0,\ldots M_{b}$,
and $m,n=1,\ldots2M_{f}$ will be included implicitly for repeated
indices.

There are the following operator correspondences for bosons \citep{Glauber_1963_P-Rep,Cahill_Glauber_OrderedExpansion_1969,Cahill_Galuber_1969_Density_operators,drummond2014quantum}:
\begin{eqnarray}
\hat{a}_{i}\hat{a}_{j}^{\dagger}\hat{\Lambda} & = & \left(\partial_{j}+\alpha_{j}^{*}\right)\alpha_{i}\hat{\Lambda}\nonumber \\
\hat{\Lambda}\hat{a}_{i}\hat{a}_{j}^{\dagger} & = & \left(\partial_{i}^{*}+\alpha_{i}\right)\alpha_{j}^{*}\hat{\Lambda}.\label{eq:identities}
\end{eqnarray}
Using an anti-normally ordered product notation for Bose operators,
$\hat{\mathcal{B}}_{ij}=\hat{a}_{i}\hat{a}_{j}^{\dagger}$, these
generally applicable identities can be written in terms of differential
operators $\mathcal{B}_{ij}$ and $\bar{\mathcal{B}}_{ij}$ as:
\begin{align}
\hat{\mathcal{B}}_{ij}\hat{\Lambda} & =\mathcal{B}_{ij}\hat{\Lambda}\nonumber \\
\hat{\Lambda}\hat{\mathcal{B}}_{ij} & =\bar{\mathcal{B}}_{ij}\hat{\Lambda}.
\end{align}

Such results also hold for fermions, except that only quadratic and
quartic fermionic terms occur, so that there is no need to treat odd
numbers of operators. We introduce $\partial_{nm}\equiv\partial/\partial\zeta_{mn}$,
where $\partial_{nm}$ is a structured derivative for antisymmetric
matrices such that $\partial_{nm}\xi_{m'n'}=\delta_{nn'}\delta_{mm'}-\delta_{nm'}\delta_{mn'}$,
and $\zeta_{mn}$ is the fermionic phase-space matrix from Eq (\ref{eq:MajOpX}).
Defining $\underline{\underline{\xi}}^{\pm}=\underline{\underline{\xi}}\pm i\underline{\underline{I}}$
, the following identities are known \citep{joseph2018phase}:
\begin{eqnarray}
\hat{\gamma}_{m}\hat{\gamma}_{n}\hat{\Lambda} & = & i\left(\xi_{mm'}^{-}\xi_{n'n}^{+}\partial_{m'n'}-\xi_{mn}^{+}\right)\hat{\Lambda}\nonumber \\
\hat{\Lambda}\hat{\gamma}_{m}\hat{\gamma}_{n} & = & i\left(\xi_{mm'}^{+}\xi_{n'n}^{-}\partial_{m'n'}-\xi_{mn}^{+}\right)\hat{\Lambda}.\label{eq:FermiIdentities}
\end{eqnarray}
Introducing quadratic Fermi operators $\mathcal{\hat{F}}_{mn}=\hat{\gamma}_{m}\hat{\gamma}_{n}$
and their corresponding differential operators $\mathcal{F}_{mn}$
and $\bar{\mathcal{F}}_{mn}$, one can write this as 
\begin{eqnarray}
\mathcal{\hat{F}}_{mn}\hat{\Lambda} & = & \mathcal{F}_{mn}\hat{\Lambda}\nonumber \\
\hat{\Lambda}\mathcal{\hat{F}}_{mn} & = & \bar{\mathcal{F}}_{mn}\hat{\Lambda}.
\end{eqnarray}

Q-function evolution equations are obtained by using these operator
identities to change Hilbert space operators acting on $\hat{\rho}$
to differential operators acting $\hat{\Lambda}$, and hence on $Q$.
To obtain operator product identities for quartic terms, one uses
the fact that the mode operators commute with the c-number terms,
so that the operator closest to the kernel $\hat{\Lambda}$ always
generates a differential term that is furthest from $\hat{\Lambda}$.
In the fermionic case one must choose a defined index ordering, for
example $m<n$, and use antisymmetry with $\xi_{nm}=-\xi_{mn}$, so
that there are only independent variables in the phase-space.

As a typical example, one obtains: 
\begin{align}
\hat{\mathcal{B}}_{ij}\hat{\mathcal{B}}_{kl}\hat{\Lambda} & =\mathcal{B}_{kl}\mathcal{B}_{ij}\hat{\Lambda}\nonumber \\
\hat{\Lambda}\hat{\mathcal{B}}_{ij}\hat{\mathcal{B}}_{kl} & =\bar{\mathcal{B}}_{ij}\bar{\mathcal{B}}_{kl}\hat{\Lambda}.\label{eq:product-identities}
\end{align}

Exact Q-functions are known for a number of special cases, including
all gaussian states. For brevity we focus on bosonic examples, as
fermionic cases are treated elsewhere \citep{Corney_PD_JPA_2006_GR_fermions,Corney_PD_PRB_2006_GPSR_fermions,FermiQ}.
A noninteracting multi-mode vacuum state, $\left|\bm{0}\right\rangle $,
and more generally a coherent state $\left|\bm{\alpha}_{0}\right\rangle _{c}$,
where $\bm{\alpha}_{0}\equiv\bm{0}$ in the vacuum state, has the
Q-function
\begin{equation}
Q^{\alpha}\left(\bm{\alpha}\right)=\frac{1}{\pi^{M}}\exp\left(-\left|\bm{\alpha}-\bm{\alpha}_{0}\right|^{2}\right)\,.
\end{equation}
This has a well known interpretation \citep{arthurs1965bstj,Leonhardt:1993}.
If one makes a simultaneous measurement of two orthogonal quadratures,
which is possible using a beam-splitter, then $Q\left(\bm{\alpha}\right)$
is the probability of a simultaneous measurement of quadratures $q$
and $p$, where $\alpha=q+ip$. This is also the result of an amplified
measurement \citep{leonhardt1993simultaneous}.

Any number state $\left|\Psi\right\rangle =\left|\bm{n}\right\rangle $
has a representation as:
\begin{align}
Q^{\alpha}\left(\bm{\alpha}\right) & =\frac{1}{\pi^{M}}e^{-\left|\bm{\alpha}\right|^{2}}\prod_{i}\left[\sum_{n_{i}}\frac{\left|\alpha_{i}\right|^{2n_{i}}}{n_{i}!}\right].
\end{align}
A free-particle thermal state with mean particle number $\bm{n}^{th}$
has a Gaussian Q-function given by:
\begin{equation}
Q^{\alpha}\left(\bm{\alpha}\right)=\prod_{i}\frac{1}{\pi\left(1+n_{i}^{th}\right)}e^{-\left|\alpha_{i}\right|^{2}/\left(1+n_{i}^{th}\right)}.
\end{equation}

\subsection{Quantum field dynamics}

To understand dynamics, we consider a time-dependent multi-mode Hamiltonian
with quartic, cubic, quadratic and linear terms. This Hamiltonian
is then expanded using mode operators. In the present treatment, although
number conserving nonlinear bosonic terms like $\hat{a}^{\dagger2}\hat{a}^{2}$
are included, we exclude non-number-conserving cubic or quartic terms
in bosonic operators like $\hat{a}^{4}$. This removes self-interacting
real scalar bosonic fields and related Hamiltonians in the standard
model. 

The models covered are applicable to many experimentally tested fundamental
as well as common effective Hamiltonians, including QED, Bose and
Fermi Hubbard models, the Ising model, parametric amplifiers and quantum
circuits. The QCD interaction Hamiltonian has a quartic self-interaction
term for Yang-Mills bosons \citep{hooft1971renormalization}. This
would require a Q-function involving Gaussian bosonic operators \citep{Corney2003}
which is not treated here.

A simple example is the QED interaction Hamiltonian, of form $\hat{\psi}_{i}\hat{A}^{\rho}\hat{\psi}_{j}$,
where $\hat{\bm{\psi}}$ is a relativistic Fermi field and $\hat{\bm{A}}$
is the electromagnetic potential. After expanding in annihilation
and creation operators, this leads to terms of the form $\hat{\mathcal{B}}_{ij}\hat{\mathcal{F}}_{mn}$
in our notation. Including all the presently allowed combinations
of fermionic and bosonic fields gives:
\begin{align}
\hat{H}\left(t\right) & =\hat{H}_{B}\left(t\right)+\hat{H}_{FB}\left(t\right)+\hat{H}_{F}\left(t\right),\label{eq:General hermitian}
\end{align}
where the different types of Hamiltonian are:
\begin{align}
\hat{H}_{B}\left(t\right) & \equiv\frac{\hbar}{2}g_{ijkl}^{B}\left(t\right)\hat{\mathcal{B}}_{ij}\hat{\mathcal{B}}_{kl}\nonumber \\
\hat{H}_{FB}\left(t\right) & \equiv\frac{\hbar}{2}g_{ijmn}^{FB}\left(t\right)\hat{\mathcal{B}}_{ij}\hat{\mathcal{F}}_{mn}\nonumber \\
\hat{H}_{F}\left(t\right) & \equiv\frac{\hbar}{2}g_{mnop}^{F}\left(t\right)\hat{\mathcal{F}}_{mn}\hat{\mathcal{F}}_{op}\,.
\end{align}

 While formally quartic, these Hamiltonians include linear, quadratic
and cubic bosonic Hamiltonians as well, through the bosonic terms
that involve $\hat{a}_{0}$. The quadratic case includes all bosonic
and fermionic free-field Hamiltonians, both relativistic and non-relativistic,
because we have defined $\hat{\mathcal{B}}_{00}=\hat{a}_{0}\hat{a}_{0}^{\dagger}=\hat{1}$.
The time argument of $g\left(t\right)$ is not written explicitly
from now on, but it is understood to apply. Renormalization is carried
out through cutoff dependent coupling constants.

As $\hat{H}$ is hermitian, the following constraints must be imposed
on the coupling matrices:
\begin{align}
g_{ijkl}^{B} & =g_{lkji}^{B*}\nonumber \\
g_{ijmn}^{FB} & =g_{jinm}^{FB*}\nonumber \\
g_{mnop}^{F} & =g_{ponm}^{F*}\,.
\end{align}
Without loss of generality, we also assume a permutation symmetry
with
\begin{align}
g_{ijkl}^{B} & =g_{klij}^{B}\nonumber \\
g_{mnop}^{F} & =g_{opmn}^{F}.
\end{align}

Cubic bosonic Hamiltonians such as $\hat{a}_{0}\hat{a}_{j}\hat{a}_{k}^{\dagger}\hat{a}_{l}^{\dagger}=\hat{a}_{j}\hat{a}_{k}^{\dagger}\hat{a}_{l}^{\dagger}$
are also of this general form. These describe parametric couplings
found in quantum science, and are used in low-noise amplifiers for
quantum measurements \citep{Drummond2004_book}. While not all quantum
field Hamiltonians in the standard model can be analyzed without further
extensions to the representation, the class of Hamiltonians that is
covered is both very large and applicable to much of current quantum
physics. 

Quantum dynamics is described by the Schr\"odinger equation, which
is
\begin{equation}
i\hbar\frac{d\hat{\rho}}{dt}=\left[\hat{H},\hat{\rho}\right].
\end{equation}
The corresponding dynamical evolution of the Q-function for unitary
evolution is given by
\begin{align}
\frac{dQ}{dt} & =\frac{i}{\hbar}Tr\left\{ \left[\hat{H},\hat{\Lambda}\left(\bm{\alpha}\right)\right]\hat{\rho}\right\} .
\end{align}

After implementing the mappings given above, one obtains three types
of differential operator acting on Q:
\begin{equation}
\frac{dQ}{dt}=\left[\mathcal{L}_{B}+\mathcal{L}_{FB}+\mathcal{L}_{F}\right]Q.
\end{equation}
The identities of Eq (\ref{eq:identities}) and (\ref{eq:product-identities})
give for boson-boson coupling, found in the Bose-Hubbard models:
\begin{align}
\mathcal{L}_{B} & =\frac{i}{2}g_{ijkl}^{B}\left[\left(\partial_{l}+\alpha_{l}^{*}\right)\alpha_{k}\left(\partial_{j}+\alpha_{j}^{*}\right)\alpha_{i}\right.\nonumber \\
 & \left.-\left(\partial_{i}^{*}+\alpha_{i}\right)\alpha_{j}^{*}\left(\partial_{k}^{*}+\alpha_{k}\right)\alpha_{l}^{*}\right].\label{eq:BB}
\end{align}
For boson-fermion coupling, one obtains from Eq (\ref{eq:FermiIdentities}):
\begin{align}
\mathcal{L}_{FB} & =\frac{-1}{2}g_{ijmn}^{FB}\left[\left(\partial_{j}+\alpha_{j}^{*}\right)\alpha_{i}\left(\xi_{mm'}^{-}\xi_{n'n}^{+}\partial_{m'n'}-\xi_{mn}^{+}\right)\right.\nonumber \\
 & \left.-\left(\partial_{i}^{*}+\alpha_{i}\right)\alpha_{j}^{*}\left(\xi_{mm'}^{+}\xi_{n'n}^{-}\partial_{m'n'}-\xi_{mn}^{+}\right)\right].\label{eq:BF}
\end{align}
The case of fermion-fermion coupling, as in the Fermi-Hubbard model,
gives:
\begin{align*}
\mathcal{L}_{F} & =\frac{-i}{2}g_{mnop}^{F}\left[\left(\xi_{oo'}^{-}\xi_{p'p}^{+}\partial_{o'p'}-\xi_{op}^{+}\right)\times\right.\\
 & \times\left(\xi_{mm'}^{-}\xi_{n'n}^{+}\partial_{m'n'}-\xi_{mn}^{+}\right)\\
 & \left.-\left(\xi_{mm'}^{+}\xi_{n'n}^{-}\partial_{m'n'}-\xi_{mn}^{+}\right)\left(\xi_{oo'}^{+}\xi_{p'p}^{-}\partial_{o'p'}-\xi_{op}^{+}\right)\right].
\end{align*}

The antisymmetry of $\xi_{mn}$ means that one must use the identity
that $\partial_{nm}=-\partial_{mn}$ in order to restrict summations
of derivatives so that only independent fermionic variables are summed
over, with $m<n$. One can include decoherence and reservoirs by adding
them to the Hamiltonian. Such reservoirs can be included in the dynamical
equations, thus enlarging both the Hilbert space and the phase space
dimension. As a result, unitary evolution is not a limitation. 

Next, define an extended vector $\vec{\alpha}=\alpha^{\mu}$, where
$\alpha^{j}=\alpha_{j}$, $\alpha^{j+M_{b}}=\alpha_{j}^{*}$, $\partial_{j+M_{b}}=\partial_{j}^{*}$,
which includes amplitudes and conjugates. If fermions are present,
one must take account of matrix antisymmetry, to include the fermionic
variables. To treat this, for indices $\rho>2M_{b}$, we define extended
indices $\rho(m,n)$ so that $\alpha^{\rho(m,n)}=\xi_{mn}$, where
$m<n$ and $\rho(m,n)=2M_{b}+m+(n-1)(n-2)/2$. 

This gives a total index range of $M=2M_{b}+M_{f}\left(2M_{f}-1\right)$.
Using an implicit Einstein summation convention, over $\mu,\nu=1,\ldots M$,
and noting that constant terms cancel due to the conservation of probability,
one obtains a complex time-symmetric Fokker-Planck equation (TFPE)
where:
\begin{align}
\frac{dQ}{dt} & =\left[\mathcal{L}_{B}+\mathcal{L}_{FB}+\mathcal{L}_{F}\right]Q\nonumber \\
 & =\left[-\frac{\partial}{\partial\alpha^{\nu}}A_{\alpha}^{\nu}\left(\vec{\alpha}\right)+\frac{1}{2}\frac{\partial}{\partial\alpha^{\mu}}\frac{\partial}{\partial\alpha^{\nu}}D_{\alpha}^{\mu\nu}\left(\vec{\alpha}\right)\right]Q.
\end{align}

From Eq (\ref{eq:BB}) and (\ref{eq:identities}), the diffusion term
for the bosonic case $\mathcal{L}_{B}$, with $1\le l,j\le M$ is:
\begin{align}
D_{\alpha}^{lj}\left(\vec{\alpha},t\right) & =i\sum_{i,k=0}^{M}g_{ijkl}^{B}\alpha_{i}\alpha_{k}\,\,.\label{eq:GeneralDiffusionTerm}
\end{align}
Letting $l',j'\equiv l+M,j+M$, one sees that 
\begin{equation}
D_{\alpha}^{lj}=D_{\alpha}^{l'j'*},\label{eq:Conjugate diffusion}
\end{equation}
and for unitary evolution there are no cross-terms $D_{\alpha}^{lj'}$.
Generally, the second-order coefficient $D_{\alpha}^{\mu\nu}\left(\vec{\alpha}\right)$
depends on the extended phase-space location $\vec{\alpha}$. In cases
of purely quadratic Hamiltonians, the diffusion is either zero or
constant in phase-space.

\subsection{Traceless diffusion and time-reversibility}

For unitary quantum evolution, the diffusion matrix is divided into
two parts, one positive definite and one negative definite, corresponding
to diffusion in the forward and backward time directions respectively.
To prove this, we first show that the corresponding Q-function time-evolution
has a TFPE with a traceless diffusion matrix. That is, unlike standard
diffusion equations, we will prove that the Q-function dynamical equation
has an equal weight of positive and negative diagonal diffusion terms. 

For length reasons, just the Bose-Bose and Bose-Fermi cases are analyzed
here. Proof that the traceless property holds in the Fermi-Fermi case
is treated elsewhere. To map Hilbert space time-evolution to phase-space
time evolution, the operator identities are utilized.  In the boson-boson
case of Eq (\ref{eq:GeneralDiffusionTerm}), terms with non-zero $l$
and $j$ indices generate second-order derivative terms which give
the diffusion matrix. If terms that multiply this have $k=i=0$ the
diffusion is constant, but otherwise it depends on the phase-space
coordinate $\vec{\alpha}$. Diagonal second order terms are obtained
when two derivatives act on the same mode.

The $k-th$ diagonal diffusion term in complex variables $\partial_{k}$
comes from from identities involving $i\hat{H}_{B}\hat{\Lambda}$
with $0<k=l\le M$ , given in Eq (\ref{eq:GeneralDiffusionTerm}).
The diagonal term in $\partial_{k}$ is accompanied by the hermitian
conjugate term derived from the reverse ordering, of form $-i\hat{\Lambda}\hat{H}_{B}$,
so that $\mathcal{L}_{B}$ is real overall. This allows the introduction
of real quadrature variables $q_{j},p_{j}$, defined such that for
$\mu=j\le M$:
\begin{equation}
\alpha_{j}=q_{j}+ip_{j}.\label{eq:real-quadratures}
\end{equation}
Hence, the derivative terms in real variables are:
\begin{equation}
\frac{\partial}{\partial\alpha_{j}}=\frac{1}{2}\left[\frac{\partial}{\partial q_{j}}-i\frac{\partial}{\partial p_{j}}\right].
\end{equation}
Defining $q^{j}=q_{j}$, $q^{j+M}=p_{j}$ gives an extended $2M$
dimensional real vector, which is written with a superscript as $q^{\mu}$.
Including the conjugate term from Eq (\ref{eq:Conjugate diffusion}),
and making this transformation, the time-symmetric Fokker-Planck equation
is:
\begin{equation}
\frac{dQ}{dt}=\left[-\frac{\partial}{\partial q^{\nu}}A_{q}^{\nu}\left(\vec{q},t\right)+\frac{1}{2}\frac{\partial}{\partial q^{\mu}}\frac{\partial}{\partial q^{\nu}}D_{q}^{\mu\nu}\left(\vec{q},t\right)\right]Q\,,
\end{equation}
where the diagonal diffusion term in real variables is:
\begin{equation}
D_{q}^{jj}=-D_{q}^{j'j'}=\frac{1}{2}Re\left(D_{\alpha}^{jj}\right).
\end{equation}

Here $j'=j+M$, and as a result, on summing the diagonal terms, the
diffusion matrix with real variables is traceless, i.e., $Tr\left[\bm{D}\right]=0.$
Given this analysis, the traceless property applies to a general class
of quadratic, cubic and quartic Hamiltonians. There can also be variables
with zero diffusion, which are deterministic and hence also have traceless
diffusion.

Q-function dynamical equations were investigated previously in special
cases, including the anharmonic oscillator \citep{Milburn:1986} and
the Dicke model \citep{altland2012quantum}. The zero-trace result
given above is generally valid for Bose and Fermi quantum fields,
and is generic to second-order Q-function unitary evolution. This
is proved above for Bose-Bose coupled fields. From Eq (\ref{eq:BF}),
Bose-Fermi couplings can never give rise to diagonal terms. The proof
of traceless diffusion in the Fermi-Fermi case is given elsewhere. 

Traceless diffusion is preserved under both uniform rescaling and
orthogonal rotations: $\bm{\phi}=\bm{O}\bm{X}$, of the real quadrature
coordinates. Since the diffusion matrix of a real TFPE is real and
symmetric, it can always be transformed into a diagonal form in the
new variables $\bm{\phi}$, using orthogonal rotations. As a result,
the transformed phase-space variables can be classified into three
groups, having positive, zero or negative diffusion, with the equation:
\begin{align}
\frac{dQ\left(\bm{\phi}\right)}{dt} & =\mathcal{L}Q\left(\bm{\phi}\right)\label{eq:quadrature-Q-TFPE}\\
 & =\partial_{\mu}\left[-A^{\mu}\left(\bm{\phi},t\right)+\frac{1}{2}\partial_{\mu}D^{\mu\mu}\left(\bm{\phi},t\right)\right]Q\left(\bm{\phi}\right).\nonumber 
\end{align}

Here, $\partial_{\mu}\equiv\partial/\partial\phi^{\mu}$. We focus
in this paper on the Bose-Bose case, and assume that there are no
zero diffusion variables, since these correspond to the trivial case
of free fields. The orthogonal rotation can always chosen for convenience
so that it results in a traceless diagonal diffusion with $D^{\mu}>0$
for $\mu\le M/2$ and $D^{\mu}<0$ for for $\mu>M$. This generates
a characteristic structure which is universal for unitary evolution
with Hamiltonians of this form. 

The above result shows that the phase-space vector $\bm{\phi}$ can
be subdivided into two complementary pairs so that $\bm{\phi}=\left(\bm{x},\bm{y}\right)$,
where the $\bm{x}$ variables have a positive definite diffusion,
and the $\bm{y}$ variables have a negative definite diffusion. The
diffusion matrix is \textbf{not} the positive-definite type found
in classical diffusion processes.  Hence, a different approach to
simulation is necessary, via diffusion in a higher space-time dimension,
as explained below.

\subsection{Constant diffusion nonlinear cases}

If the Hamiltonian has only quadratic terms, the diffusion terms are
either zero or constant in phase-space. In cases of a nonlinear, number
preserving quartic interaction there is a nonlinear transformation
that also gives constant diffusion for the most common form of \emph{bosonic
nonlinear} coupling, namely density-density coupling. The result is
an alternative definition of the transformed variable $\bm{\phi}$.
Since we wish to focus on constant diffusion cases, the proof of a
transformation to a constant diffusion TFPE is given in this subsection.
We will show that in this nonlinear case there is a TFPE with constant
diffusion, independent of $\bm{\phi}$, as well as being traceless
and diagonal. 

This type of physics is found in the Bose-Hubbard model and other
bosonic quantum field theories \citep{Drummond_1987_JOptSocAmB,zee2010quantum}. 

On a lattice, consider a quartic Hamiltonian of form:
\begin{equation}
\hat{H}_{B}=\hbar\sum_{ij}^{M_{b}}\left[\omega_{ij}\hat{a}_{i}^{\dagger}\hat{a}_{j}+\frac{1}{2}g_{ij}\hat{a}_{i}^{\dagger}\hat{a}_{i}\hat{a}_{j}^{\dagger}\hat{a}_{j}\right]\,.
\end{equation}
Using the identities of Eq (\ref{eq:identities}) again, the second-order
derivative terms in $\bm{\alpha}$ are:
\begin{align}
\mathcal{L}_{B}^{(2)} & =\frac{ig_{ij}}{2}\frac{\partial}{\partial\alpha_{j}}\frac{\partial}{\partial\alpha_{i}}\alpha_{j}\alpha_{i}+H.c.\,.
\end{align}
In this case one may define a mapping, $\theta_{j}=\lambda\ln(\alpha_{j})$,
where $\lambda$ is a scaling factor, so that in the new variables
the diffusion matrix $D_{ij}^{\theta}$ is constant, where:
\begin{equation}
D_{ij}^{\theta}=i\lambda^{2}g_{ij}.
\end{equation}

This transformation simplifies the time-symmetric Fokker-Planck path
integral. Path integrals for space-dependent diffusion as in Eq (\ref{eq:quadrature-Q-TFPE})
exist \citep{graham1977path}, but are more complex. If the diffusion
is constant, as in quadratic Hamiltonians, this step is unnecessary.

\begin{figure}
\centering{}\includegraphics[width=0.75\columnwidth]{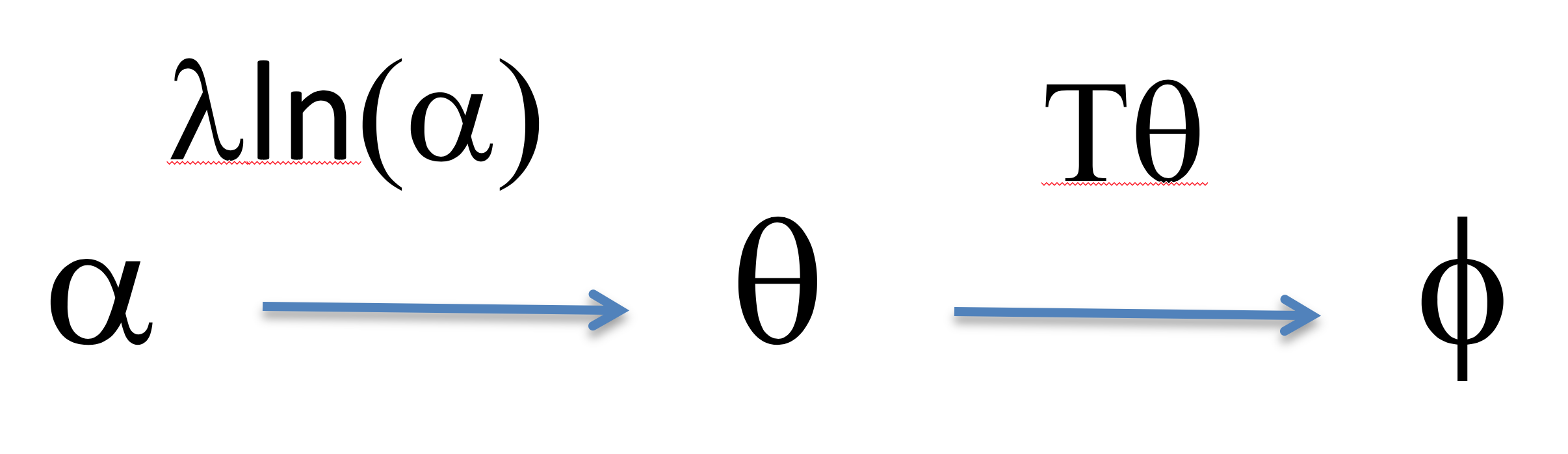}\caption{Transformations used in the phase-space. The original complex mode
amplitudes $\bm{\alpha}$ are transformed first to constant-diffusion
mode amplitudes $\bm{\theta}=\lambda\ln\left(\bm{\alpha}\right)$,
then mapped with a possibly time-dependent mapping to real quadrature
amplitudes $\bm{\phi}=\bm{T}\bm{\theta}$.\label{fig:Transformations-used-in}}
\end{figure}

We now check that the traceless property persists after making this
variable change to logarithmic variables. The Q-function is mapped
to a set of constant diffusion, complex phase-space variables $\bm{\theta}$,
as shown in Fig (\ref{fig:Transformations-used-in}), which satisfy
an equation of form:
\begin{equation}
\frac{\partial}{\partial t}Q^{\theta}=\mathcal{L}^{\theta}Q^{\theta}\,.
\end{equation}

To prove the traceless property, a second mapping is made to a real
quadrature vector, $\bm{\phi}=\left[\phi^{1},\ldots\phi^{2M}\right]$,
described by the linear transformation $\bm{\phi}=\bm{T}\bm{\theta}.$
In this constant diffusion space, there are diagonal second derivative
terms together with conjugate terms such that $D_{jj}^{\theta}=e^{-2i\eta_{j}}\left|D_{jj}^{\theta}\right|$,
where $D_{jj}^{\theta}=i\lambda^{2}g_{jj}$. The corresponding real
variables are defined in this case as:
\begin{equation}
\theta_{j}e^{i\eta_{j}}=x_{j}+iy_{j},\label{eq:xy-quadratures}
\end{equation}
where $\bm{\phi}=\left(\bm{x},\bm{y}\right)$. For a one-mode case,
the mapping transformation matrix from complex logarithmic to real
variables is
\begin{equation}
\bm{T}=\left[\begin{array}{cc}
e^{i\eta_{j}} & e^{-i\eta_{j}}\\
-ie^{i\eta_{j}} & ie^{-i\eta_{j}}
\end{array}\right].
\end{equation}

As a result, the diagonal diffusion term in $\phi$ is
\begin{equation}
D^{jj}=-D^{j'j'}=\frac{1}{2}\left|D_{jj}^{\theta}\right|,
\end{equation}
where $j'=j+M$. There is a positive diffusion in $\bm{x}$ and negative
diffusion in $\bm{y}$ as described in the previous subsection, except
with a constant diffusion matrix.

The result is a transformed Q-function, $Q=Q^{\theta}\left|\delta\bm{\theta}/\delta\bm{\phi}\right|$,
which evolves according to real differential equation. Introducing
$\partial_{\mu}\equiv\partial/\partial\phi^{\mu}$, for $\mu=1,\ldots2M$,
the time-evolution equation is a TFPE with a diagonal, constant diffusion
identical to Eq (\ref{eq:quadrature-Q-TFPE}).

The transformed diffusion matrix is traceless as shown previously,
so that $\sum D^{\mu\mu}=0\,.$ The phase-space probability $Q$ is
positive, yet since the overall diffusion term $\bm{D}$ is \emph{not}
positive definite, this is not a forward-time stochastic process \citep{Gardiner_Book_SDE}.
The form of Eq (\ref{eq:quadrature-Q-TFPE}) means that probability
is conserved by the dynamical equations, provided boundary terms vanish,
which is also required from the definition and Eq (\ref{Q-normalization}),
so that for a $2M$-dimensional real measure $d\bm{\phi}$:
\begin{equation}
\int d\bm{\phi}Q\left(\bm{\phi},t\right)=1\,.\label{eq:Conservation_from_dynamics}
\end{equation}

The Q-function obeys a second order partial differential equation.
Yet it describes a \emph{reversible} process. Positive distribution
functions in statistical mechanics commonly follow a diffusion equation
which is irreversible, owing to couplings to a reservoir. Since the
Q-function is a phase-space representation that is positive, it can
be treated and sampled in a similar way to a  probability distribution. 

\section{Time-symmetric stochastic action\label{sec:Bidirectional-stochastic-bridges}}

The Q-function for unitary evolution can be transformed to satisfy
a real partial differential equation, with a traceless diffusion term
that is \emph{not} positive-definite. The resulting initial value
problem is not well-posed unless the initial conditions have compact
support in Fourier space \citep{miranker1961well}. Hence, Green's
functions with delta-function initial conditions \citep{Gardiner1997}
leading to a forward-time stochastic differential equation are not
defined. 

The alternative approach introduced here is to use Green's functions
with both initial\emph{ }and final boundary conditions for the TFPE,
Eq (\ref{eq:quadrature-Q-TFPE}). Our goal is to derive a quantum
action principle \citep{dirac2005lagrangian} or path integral in
Feynman's terminology \citep{feynman2010quantum}, with a real action
integral. The leads to probabilistic evolution in space-time in which
one can identify continuous sample trajectories.

The traceless property of the Q-function diffusion means that the
phase space of $\bm{\phi}$ is generally divisible into two $M$-dimensional
sub-vectors, so that $\bm{\phi}=\left[\bm{x},\bm{y}\right]$. These
have the physical interpretation of complementary variables. The $\bm{x}$
fields will be called positive-time fields, with indices in the set
$T_{+}$, while the $\bm{y}$ fields will be called negative-time
fields, with indices in the set $T_{-}$, so that:
\begin{align}
\bm{D} & =\left[\begin{array}{cc}
\bm{d}^{x} & \bm{0}\\
\bm{0} & -\bm{d}^{y}
\end{array}\right],\nonumber \\
\bm{A}\left(\bm{\phi}\right) & =\left[\begin{array}{c}
\bm{a}^{x}\left(\bm{\phi}\right)\\
-\bm{a}^{y}\left(\bm{\phi}\right)
\end{array}\right].
\end{align}

The $\bm{d}^{x,y}$ matrices are assumed positive definite, where
$d_{ij}^{x}=\left\{ D^{ij}\right\} $ and $d_{ij}^{y}=\left\{ -D^{i+M,j+M}\right\} $
 for $i,j\le M$. This is a more general scenario than a strictly
diagonal diffusion, but includes it as a special case. Hence, there
are two positive-definite differential operators, $\mathcal{L}^{x,y}$,
such that $\mathcal{L}=\mathcal{L}^{x}-\mathcal{L}^{y}$, where:
\begin{align}
\mathcal{L}^{x}= & -\partial_{i}^{x}a_{i}^{x}\left(\bm{\phi}\right)+\frac{1}{2}\partial_{i}^{x}\partial_{j}^{x}d_{ij}^{x}\nonumber \\
\mathcal{L}^{y}= & -\partial_{i}^{y}a_{i}^{y}\left(\bm{\phi}\right)+\frac{1}{2}\partial_{i}^{y}\partial_{j}^{y}d_{ij}^{y}.
\end{align}
One usually solves for $Q\left(\bm{\phi},t\right)$ at a later time
$t>t_{0}$, given an initial distribution $Q_{0}\left(\bm{\phi},t_{0}\right)$.
However, not having positive-definite diffusion in $\bm{y}$ means
that one cannot use standard Green's functions or propagators to propagate
$Q$ forward in time, without requiring singular Green's functions.
One also can't propagate $Q$ purely backward in time, for the same
reason.  

\subsection{Input and output boundary values}

We will define time-symmetric propagators relative to combined boundary
value conditions in the past \emph{and} the future for the TFPE. We
introduce a notation $P\left(\bm{\phi}_{OUT}\left|\bm{\phi}_{IN}\right.\right)$
to indicate the probability density of \emph{output} event(s) $\bm{\phi}_{OUT}$
given \emph{input} event(s) $\bm{\phi}_{IN}$. This does not imply
time-ordering, and is different to the usage in quantum field theory
\citep{weinberg2015lectures} where time-ordering is implied. Both
\emph{output }and \emph{input }events may include past and future
times. 

The input boundary coordinates will be labeled $\bm{\phi}_{IN}=\left(\bm{x}_{0},\bm{y}_{f}\right)$.
This indicates a boundary at $t_{0}$ for positive-time fields $\bm{x}$,
and at $t_{f}$ for negative-time fields $\bm{y}$. The complementary
boundary $\bm{\phi}_{OUT}=\left(\bm{x}_{f},\bm{y}_{0}\right)$ is
the output of the quantum process. As explained above, the terms \emph{input}
and \emph{output} are used to indicate causality, not time-ordering.
 The joint probability of the input events in the past and future
is defined as $P\left(\bm{\phi}_{IN},\bm{t}\right)$, where $\bm{t}=\left(t_{0},t_{f}\right)$
are the times involved.

Defining $d\bm{x}$ and $d\bm{y}$ as $M$-dimensional real measures,
marginal distributions at the same time for $\bm{x}$ follow the usual
conventions where:
\begin{equation}
P_{x}\left(\bm{x},t\right)=\int P\left(\bm{\phi},t\right)d\bm{y}\,,\label{eq:marginal_x}
\end{equation}
andmarginal distributions for $\bm{y}$ are: 
\begin{equation}
P_{y}\left(\bm{y},t\right)=\int P\left(\bm{\phi},t\right)d\bm{x}\,.\label{marginal_y}
\end{equation}

In some cases, the boundary values for the input distribution $P\left(\bm{\phi}_{IN},t\right)$
are independent of each other. This implies that one can write the
joint probability of $\bm{x}_{0}$ in the past and and $\bm{y}_{f}$
in the future as a product of two independent distributions, so that
\begin{equation}
P\left(\bm{\phi}_{IN},\bm{t}\right)=P_{x}\left(\bm{x}_{0},t_{0}\right)P_{y}\left(\bm{y}_{f},t_{f}\right).
\end{equation}

\begin{figure}
\includegraphics[width=1\columnwidth]{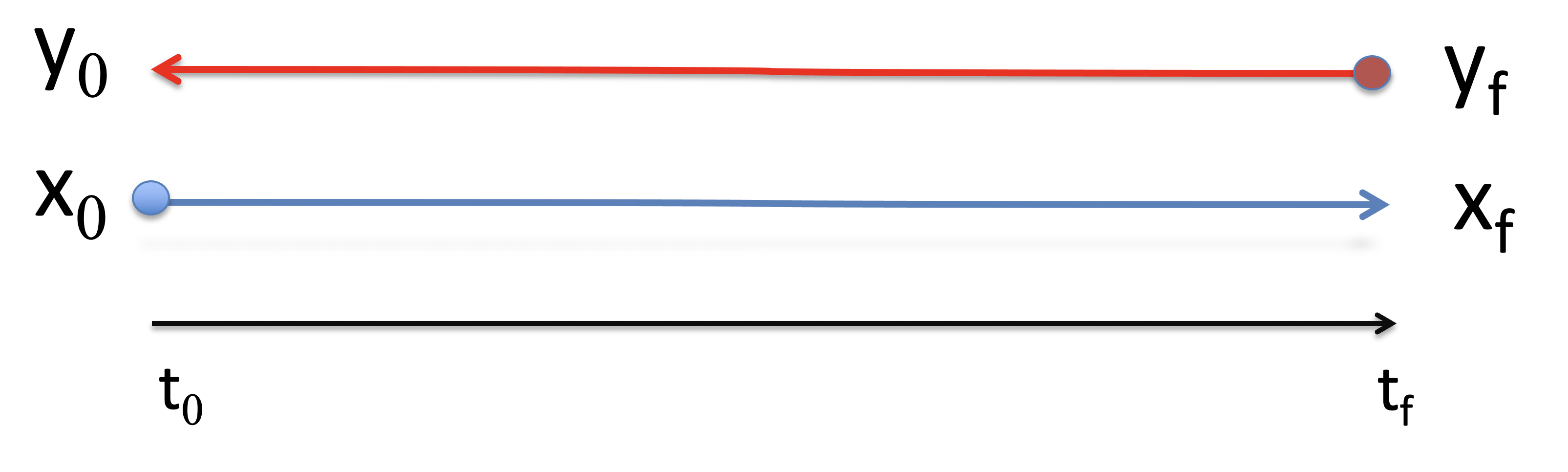}\\
\caption{Quantum fields propagating in phase space, from initial time $t_{0}$
to final time $t_{f}$. The $\bm{x}$ components propagate in the
positive time direction, while the $\bm{y}$ components propagate
in the negative time direction. The joint probability of input events
is $P\left(\bm{\phi}_{IN},\bm{t}\right)$, where $\bm{\phi}_{IN}=\left(\bm{x}_{0},\bm{y}_{f}\right)$.
\label{fig:Quantum-fields-propagating-full}}
\end{figure}

In general, there are correlations, and the boundary value distribution
$P\left(\bm{\phi}_{IN},\bm{t}\right)$ cannot be factorized. This
is the more general case that we analyze here. A graphical illustration
of this is given in Fig (\ref{fig:Quantum-fields-propagating-full}).
We use the convention that $P\left(\bm{\phi},t\right)$ is a general
field probability, while $Q\left(\bm{\phi},t\right)$ is a probability
in the restricted case which represents a quantum state.

\subsection{Time-symmetric propagator}

In contrast to the usual time-asymmetric propagator definitions \citep{graham1977path},
we now define a time-symmetric propagator (TSP). This is a function
$G\left(\bm{\phi},t\left|\bm{\phi}_{IN},\bm{t}\right.\right)$, relative
to inputs $\bm{\phi}_{IN}=\left[\bm{x}{}_{0},\bm{y}{}_{f}\right]$
and times $\bm{t}=\left[t_{0},t_{f}\right]$, where $t_{0}\le t\le t_{f}$.
This is defined to satisfy the Q-function evolution equation with
both initial \emph{and} final delta-function boundary values:
\begin{equation}
\frac{\partial}{\partial t}G\left(\bm{\phi},t\left|\bm{\phi}_{IN},\bm{t}\right.\right)=\mathcal{L}G\left(\bm{\phi},t\left|\bm{\phi}_{IN},\bm{t}\right.\right).\label{eq:TSP-equation}
\end{equation}

Dropping the conditional arguments for brevity, the general equation
of motion for the TSP is
\begin{align}
\frac{dG\left(\bm{\phi},t\right)}{dt} & =\left[-\partial_{\mu}A^{\mu}\left(\bm{\phi}\right)+\frac{1}{2}\partial_{\mu}\partial_{\nu}D^{\mu\nu}\left(\bm{\phi}\right)\right]G\left(\bm{\phi},t\right)\nonumber \\
 & =\left[\mathcal{L}^{x}-\mathcal{L}^{y}\right]G\left(\bm{\phi},t\right),\label{eq:TSP-evolution-equation}
\end{align}
with the property that $G\left(\bm{\phi},t\right)$ is positive and
normalized so that
\begin{equation}
\int G\left(\bm{\phi},t\right)d\bm{\phi}=1.
\end{equation}

The corresponding marginal distributions are given by:
\begin{align}
G_{x}\left(\bm{x},t\left|\bm{\phi}_{IN},\bm{t}\right.\right) & =\int d\bm{y}G\left(\bm{\phi},t\left|\bm{\phi}_{IN},\bm{t}\right.\right)\nonumber \\
G_{y}\left(\bm{y},t\left|\bm{\phi}_{IN},\bm{t}\right.\right) & =\int d\bm{x}G\left(\bm{\phi},t\left|\bm{\phi}_{IN},\bm{t}\right.\right).
\end{align}
The initial and final boundary conditions are delta-correlated marginal
distributions, where $\delta\left(\bm{x}\right)$ is an $M-$dimensional
real Dirac delta function, so that:
\begin{align}
G_{x}\left(\bm{x},t_{0}\left|\bm{\phi}_{IN},\bm{t}\right.\right) & =\delta\left(\bm{x}-\bm{x}{}_{0}\right)\nonumber \\
{G_{y}\left(\bm{y},t_{f}\left|\bm{\phi}_{IN},\bm{t}\right.\right)} & {=\delta\left(\bm{y}-\bm{y}{}_{f}\right).}\label{eq:marginal-delta-TSP}
\end{align}
This has the interpretation that, given input boundary values $\bm{\phi}_{IN}=\left[\bm{x}{}_{0},\bm{y}{}_{f}\right]$
at $t_{0},t_{f}$, the probability density of obtaining the field
values $\bm{\phi}=\left[\bm{x},\bm{y}\right]$ at $t$ is $G\left(\bm{\phi},t\left|\bm{\phi}_{IN},\bm{t}\right.\right)$. 

Since $\mathcal{L}$ is a linear operator, any linear combination
or integral of propagators also satisfies the propagator time-evolution
equation, (\ref{eq:TSP-equation}). As a result, provided that $G\left(\bm{\phi},t_{0}\left|\bm{\phi}_{IN},\bm{t}\right.\right)$
can be integrated with respect to a nonlocal joint distribution $P\left(\bm{\phi}_{IN}\right)$
to obtain a Q-function at $t_{0}$, this gives a Q-function solution
for all times:
\begin{equation}
Q\left(\bm{\phi},t\right)=\int d\bm{\phi}_{IN}G\left(\bm{\phi},t\left|\bm{\phi}_{IN},\bm{t}\right.\right)P\left(\bm{\phi}_{IN}\right)\label{eq:Q-function-from-TSP}
\end{equation}
When integrated over $\bm{x}_{f}$, $P\left(\bm{\phi}_{IN}\right)$
is a $Q$-function marginal distribution in $\bm{y}_{0}$ at time
$t_{0}$, while if integrated $\bm{y}_{0}$ it is a $Q$-function
marginal distribution in $\bm{x}_{f}$ at time $t_{f}$. 

Having a nonlocal distribution $P\left(\bm{\phi}_{IN}\right)$ for
the Q-function is not guaranteed in every case. A Q-function that
is defined initially does not  guarantee that $P\left(\bm{\phi}_{IN}\right)$
always exists. There may be Hamiltonians and states without any corresponding
nonlocal distribution. In these cases we conjecture that one must
define additional constraints or conditional boundaries, which are
outside the scope of this paper.

\subsection{Time-symmetric stochastic differential equation}

To obtain a solution for the TSP, we introduce time-symmetric stochastic
differential equations to define a set of stochastic paths. Subsequently,
we will demonstrate that an integral over paths is a solution for
the TSP, and satisfies the Q-function differential equation.

Time symmetric stochastic path equations are stochastic equations
with both future and past boundary conditions \citep{nualart1991boundary}.
These will be written in an intuitive form similar to a forward-backward
stochastic differential equation \citep{ma1994solving}. The time-symmetric
stochastic differential equation or TSSDE is defined to have the following
structure, expressed as an integral:
\begin{align}
\bm{x}(t) & =\bm{x}_{0}+\int_{t_{0}}^{t}\bm{a}^{x}\left(\bm{\phi}\left(t'\right)\right)dt'+\int_{t_{0}}^{t}d\bm{w}^{x}\nonumber \\
\bm{y}(t) & =\bm{y}_{f}+\int_{t}^{t_{f}}\bm{a}^{y}\left(\bm{\phi}\left(t'\right)\right)dt'+\int_{t}^{t_{f}}d\bm{w}^{y}\,.\label{eq:Time-symmetricSDE}
\end{align}
 The two quadrature fields are propagated in the positive and negative
time directions respectively, while the independent real Gaussian
noise terms $d\bm{w}$ are correlated over short times so that, over
a small interval $dt$:
\begin{align}
\left\langle dw_{i}^{x}\left(t\right)dw_{j}^{x}\left(t\right)\right\rangle  & =d_{ij}^{x}dt\,\nonumber \\
\left\langle dw_{i}^{y}\left(t\right)dw_{j}^{y}\left(t\right)\right\rangle  & =d_{ij}^{y}dt\,\nonumber \\
\left\langle dw_{i}^{x}\left(t\right)dw_{j}^{y}\left(t\right)\right\rangle  & =0\,
\end{align}

These equations unify two important features: time symmetry and randomness.
Similar equations occur in stochastic control theory, and there is
literature on their properties \citep{ma1994solving}, in a modified
form. In our equations the boundary values in $\bm{x}_{0}$ and $\bm{y}_{f}$
are fixed rather than conditional on outputs. There is no third field
as in some control theory equations. Analytic equations like this
may be used to develop a stochastic perturbation theory \citep{Chaturvedi1999}
for quantum fields, and in some cases may have numerical iterative
solutions. 

While these provide insight into time symmetry, they clearly cannot
be treated using conventional algorithms for ordinary stochastic differential
equations. This can be recognized by attempting to write the equations
as ordinary forward time stochastic differential equations. We may
define $\bar{\bm{y}}\left(t\right)$ as a time-reversed copy of $\bm{y}\left(t\right)$,
i.e., let $t_{-}=t_{0}+t_{f}-t$, and
\begin{equation}
\bar{\bm{y}}\left(t\right)=\text{\ensuremath{\bm{y}\left(t_{-}\right)}}\,.
\end{equation}
The stochastic differential equation that results is:
\begin{align}
d\bm{x} & =\bm{a}^{x}\left(\bm{x}\left(t\right),\bar{\bm{y}}\left(t_{-}\right),t\right)dt+d\bm{w}^{x}\nonumber \\
d\bar{\bm{y}} & =-\bm{a}^{y}\left(\bm{x}\left(t_{-}\right),\bar{\bm{y}}\left(t\right),t_{-}\right)dt-d\bm{w}^{y}\,.
\end{align}
Here, $\bm{x}\left(t_{0}\right)=\bm{x}_{0}$ and $\bar{\bm{y}}\left(t_{0}\right)=\bm{y}_{f}$
are now ``initial'' conditions, but with the $y$ coordinate replaced
by $\bar{y}$. A time-symmetric stochastic differential equation corresponds
to stochastic propagation with drift terms having fields at different
times. This non-locality in time prevents the direct use of standard
local-time algorithms for solving the equations.

This behavior is not surprising, physically. If the fields had local
drift terms, they would be causal theories that satisfy Bell's theorem,
which therefore do not correspond to quantum theory. 

\subsection{Discretized TSSDE}

To analyze the stochastic equations, consider a time-symmetric stochastic
trajectory discretized for times $t_{k}=t_{0}+k\epsilon$, with $k=0,\ldots n$,
so that $t_{f}=t_{n}$ and $\bm{\phi}_{f}=\bm{\phi}_{n}$, where $\bm{\phi}_{k}\equiv\bm{\phi}\left(t_{k}\right)$.
The notation $\bm{\phi}_{kj}\equiv\left(\bm{x}_{k},\bm{y}_{j}\right)$
is used for a field with quadratures at two different times. We now
define an $n$-step path probability $\mathcal{G}\left(\left[\bm{\phi}\right]\left|\bm{\phi}_{0n}\right.\right)$
, where the input events are $\bm{\phi}_{IN}=\bm{\phi}_{0n}$.

\begin{figure}
\includegraphics[width=1\columnwidth]{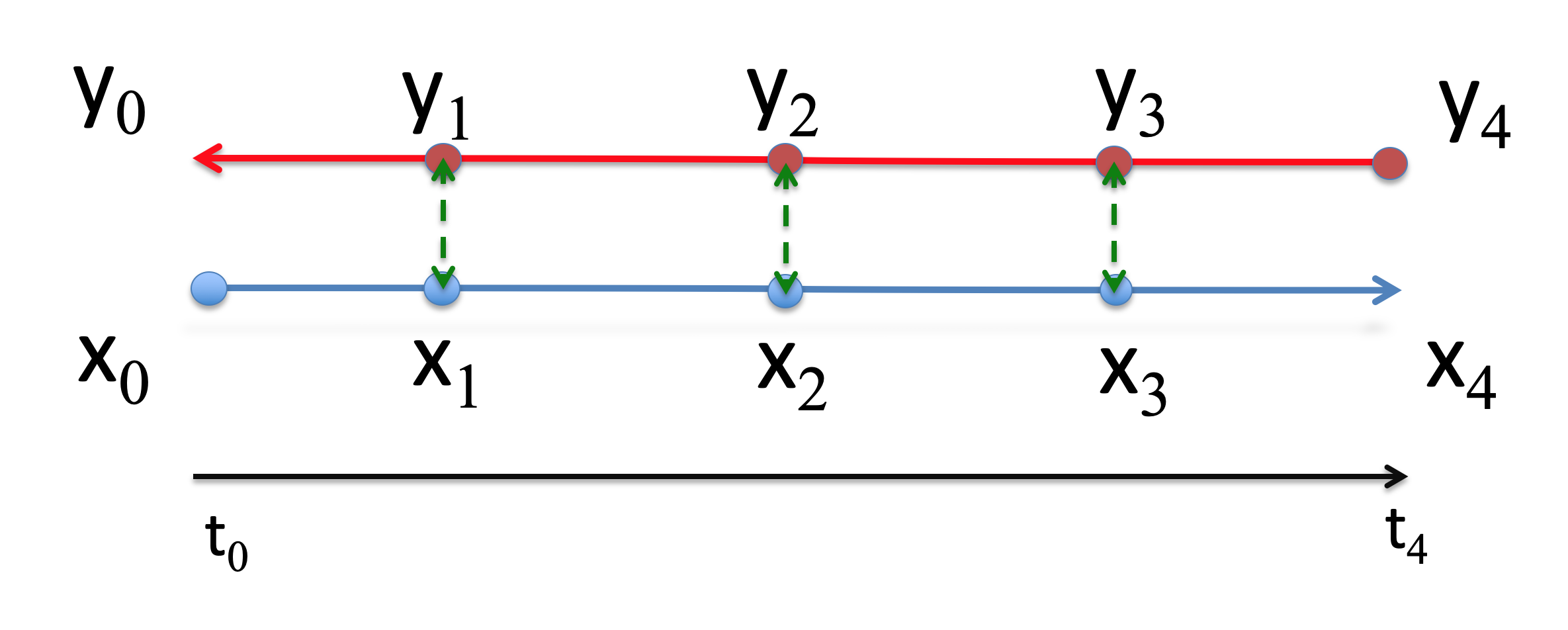}\caption{Quantum fields propagating over multiple time-intervals in phase space.
Their interactions can lead to correlations. Coupling is indicated
by green arrows.\label{fig:Quantum-field-propagating}}
\end{figure}

Here ${\left[\bm{\phi}\right]=\left[\bm{\phi}_{0},\bm{\phi}_{1},\dots\bm{\phi}_{n-1},\bm{\phi}_{n}\right]}$,
so this is the probability density of stochastic paths with points
$\left[\bm{\phi}\right]$. This is shown diagrammatically in Fig (\ref{fig:Quantum-field-propagating}
). This type of path probability is always defined relative to specific
time-symmetric inputs, $\bm{\phi}_{0n}\equiv\left(\bm{x}_{0},\bm{y}_{n}\right)$. 

To simplify results with no loss of generality, we use orthogonal
transformations and rescaling dilatations on $\bm{x}$ and $\bm{y}$,
so that each diffusion matrix is diagonal: hence $\bm{d}^{x,y}=\bm{I}d$.
The discretized equations are given by:
\begin{align}
\bm{x}_{k} & =\bm{x}_{k-1}+\bm{a}^{x}\left(\bm{\phi}_{k}^{\epsilon_{x}}\right)\epsilon+\bm{\Delta}_{k}^{x}\nonumber \\
\bm{y}_{k-1} & =\bm{y}_{k}+\bm{a}^{y}\left(\bm{\phi}_{k}^{\epsilon_{y}}\right)\epsilon+\bm{\Delta}_{k}^{y}.\label{eq:Time-symmetricSDE-1-2}
\end{align}
These have Gaussian real noises $\bm{\Delta}_{k}^{x}$ such that,
at each step in time, 
\begin{align}
\left\langle \Delta_{i}^{x}\Delta_{j}^{x}\right\rangle  & =\epsilon d\delta_{ij}\nonumber \\
\left\langle \Delta_{i}^{y}\Delta_{j}^{y}\right\rangle  & =\epsilon d\delta_{ij}\nonumber \\
\left\langle \Delta_{i}^{x}\Delta_{j}^{y}\right\rangle  & =0.
\end{align}
The time-symmetric SDE can be written in more than one equivalent
discrete form. The discretized arguments of $\bm{a}^{x,y}$ generally
are described by a $2\times2$ matrix of coefficients $s_{zk}$, where
$z=x,y$; $k=1,2$, and $0<s_{zk}<1$. These denote how the arguments
of $\bm{a}^{x,y}$ are interpolated in terms of $\bm{\phi}_{k},\bm{\phi}_{k-1}$
. The general discretized form of the drift argument is $\bm{\phi}_{k}^{s}$,
interpolated so that
\begin{equation}
\bm{\phi}_{k}^{s}=\left(s_{1}\bm{x}_{k-1}+\left(1-s_{1}\right)\bm{x}_{k},s_{2}\bm{y}_{k-1}+\left(1-s_{2}\right)\bm{y}_{k}\right).
\end{equation}
These lead to equivalent discretized TSSDEs which all give the same
stochastic process as $\delta\rightarrow0$. However, they do not
have the same path integral representations. 

There are three particular types of TSSDE discretizations that are
important here. These will be labelled forms I, II and III, depending
on the way that the discretized arguments to the drift are calculated: 

\paragraph*{TSSDE I:}

The $x$-arguments of $\bm{a}^{x,y}$ are evaluated at the earlier
time, with $y$-arguments evaluated at the later time of each step,
so $s_{z1}=1$, $s_{z2}=0$:
\begin{align}
{\bm{x}_{k}} & {=\bm{x}_{k-1}+\bm{a}^{x}\left(\bm{\phi}_{k-1,k}\right)\epsilon+\bm{\Delta}_{k}^{x}}\nonumber \\
{\bm{y}_{k-1}} & {=\bm{y}_{k}+\bm{a}^{y}\left(\bm{\phi}_{k-1,k}\right)\epsilon+\bm{\Delta}_{k}^{y}.}\label{eq:Time-symmetricSDE-1-1}
\end{align}
This is an explicit equation, as the drift argument requires no knowledge
of the solution for the next step for either variable. Such methods
are similar to the Ito theory of ordinary stochastic differential
equations \citep{Gardiner1997}.

\paragraph*{TSSDE II:}

Both arguments of $\bm{a}^{x,y}$ are evaluated at the beginning of
each step in a forwards or backwards direction respectively, so $s_{xj}=1$,
$s_{yj}=0$:
\begin{align}
{\bm{x}_{k}} & {=\bm{x}_{k-1}+\bm{a}^{x}\left(\bm{\phi}_{k-1}\right)\epsilon+\bm{\Delta}_{k}^{x}}\nonumber \\
{\bm{y}_{k-1}} & {=\bm{y}_{k}+\bm{a}^{y}\left(\bm{\phi}_{k}\right)\epsilon+\bm{\Delta}_{k}^{y}.}\label{eq:Time-symmetricSDE-1}
\end{align}
This is a quasi-explicit equation, since the drift argument for $\bm{x}$
requires a knowledge of the next step for $\bm{y}$, and vice-versa.
As a result, this has  similar properties to type I. 

\paragraph*{TSSDE III:}

In this fully symmetric form, the arguments of $\bm{a}^{x,y}$ are
evaluated at the midpoint of each step, so $s_{zj}=\frac{1}{2}$.
Defining $\bm{\bar{\phi}}_{k}=\left(\bm{\phi}_{k}+\bm{\phi}_{k-1}\right)/2$
gives:
\begin{align}
{\bm{x}_{k}} & {=\bm{x}_{k-1}+\bm{a}^{x}\left(\bm{\bar{\phi}}_{k}\right)\epsilon+\bm{\Delta}_{k}^{x}}\nonumber \\
{\bm{y}_{k-1}} & {=\bm{y}_{k}+\bm{a}^{y}\left(\bm{\bar{\phi}}_{k}\right)\epsilon+\bm{\Delta}_{k}^{y}.}\label{eq:Time-symmetricSDE-1-1-1}
\end{align}
This is an implicit algorithm \citep{drummond1991computer}, as each
drift argument requires knowledge of the solution for the next step
in both variables. 

Such techniques are often used to treat stochastic differential equations,
as the equations follow standard calculus rules. This approach is
similar to the Stratonovich theory of ordinary stochastic {differential
equations \citep{stratonovich1971probability,Drummond1990,Gardiner1997}.}
There are other discretizations possible as well, based on the general
interpolation formula above.

\subsection{Path integral and action}

To obtain a path integral representation, we generalize methods used
for forward stochastic differential equations \citep{chow2015path}.
The path probability density $\mathcal{G}\left(\left[\bm{\phi}\right]\left|\bm{\phi}_{0n},\left[\Delta\right]\right.\right)$
is conditioned both on an input $\bm{\phi}_{0n}$ and a random noise
vector, $\left[\bm{\Delta}\right]=\bm{\Delta}_{1}^{x},\bm{\Delta}_{1}^{y},\ldots\bm{\Delta}_{n}^{x},\bm{\Delta}_{n}^{y}$,
where $\bm{\Delta}_{j}=\left(\bm{\Delta}_{j}^{x},\bm{\Delta}_{j}^{y}\right)$
are independent $2M-$dimensional real Gaussian noises. This conditional
path probability is a product of Dirac delta functions, since only
one path exists for a given initial condition and noise sample:
\begin{align}
\mathcal{G}_{n}\left(\left[\bm{\phi}\right]\left|\bm{\phi}_{0n},\left[\Delta\right]\right.\right) & =\prod_{j=1}^{n}N_{j}\delta^{2M}\left(\epsilon\bm{v}_{j}-\bm{\Delta}_{j}\right).\label{eq:Delta-function-path}
\end{align}
Here we define relative velocity fields $\bm{v}=\left(\bm{v}_{j}^{x},\bm{v}_{j}^{y}\right)$
that correspond to a particular discretization, given above:
\begin{align}
\bm{v}_{j}^{x} & \equiv\frac{1}{\epsilon}\left(\bm{x}_{j}-\bm{x}_{j-1}\right)-\bm{a}^{x}\left(\bm{\phi}_{j}^{s_{x}}\right)\nonumber \\
\bm{v}_{j}^{y} & \equiv\frac{1}{\epsilon}\left(\bm{y}_{j-1}-\bm{y}_{j}\right)-\bm{a}^{y}\left(\bm{\phi}_{j}^{s_{y}}\right).\label{eq:velocity}
\end{align}

The normalization factor $N_{j}$ ensures that each term is normalized
to unity when integrated over the output quadratures, which are $\bm{x}_{j}$
and $\bm{y}_{j-1}$ respectively. Since the delta-function can be
written in the form of $\delta^{M}\left(\bm{x}_{j}-\bm{f}_{j}^{x}\right)\delta^{M}\left(\bm{y}_{j-1}-\bm{f}_{j}^{y}\right)$,
the normalization includes derivatives of $\bm{f}_{j}^{x,y}$ . The
simplest case is if the stochastic drift is defined explicitly using
the input values of the quadrature field for that transition. This
implies that $s_{x1}=1$ and $s_{y2}=0$, {which includes}
TSSDE types I and II. For types I and II, we therefore have $N_{j}=1$. 

Using delta-function integration identities, the general normalization
is
\begin{equation}
N_{j}=N_{j}^{x}N_{j}^{y}=\prod_{k}\left(1-\epsilon\frac{\partial a^{xk}\left(\bm{\phi}_{j}^{s_{x}}\right)}{\partial x_{j}^{k}}\right)\left(1-\epsilon\frac{\partial a^{yk}\left(\bm{\phi}_{j}^{s_{y}}\right)}{\partial y_{j-1}^{k}}\right).
\end{equation}
As a result, when there is an implicit drift, as in type III discretization,
the normalization is modified. 

Employing a Fourier transform representation of the delta function,
where $d\bm{k}_{j}$ is a $2M$-dimensional real measure, this becomes:
\begin{align}
\mathcal{G}\left(\left[\bm{\phi}\right]\left|\bm{\phi}_{0n},\left[\Delta\right]\right.\right) & =\prod_{j=1}^{n}\int\frac{N_{j}d\bm{k}_{j}}{\left(2\pi\right)^{2M}}e^{-i\bm{k}_{j}\left(\bm{v}_{j}\delta-\bm{\Delta}_{j}\right)}.
\end{align}
For the $2M$ real Gaussian noises $\bm{\Delta}_{k}$, at each step
in time, the probabilities of $\bm{\Delta}_{k}$ are:
\begin{equation}
P\left(\bm{\Delta}_{k}\right)=\frac{1}{\left(2\pi\epsilon d\right)^{M}}e^{-\left|\bm{\Delta}_{k}\right|^{2}/\left(2\epsilon d\right)}.
\end{equation}
Hence the path probability conditioned on the inputs only can be written
as a weighted integral over the step $\Delta$. This has the form,
after $n$ steps, of a product of $n$ successive one-step path probabilities:
\begin{align}
\mathcal{G}\left(\left[\bm{\phi}\right]\left|\bm{\phi}_{0n}\right.\right) & =\prod_{j=1}^{n}\mathcal{G}\left(\bm{\phi}_{j-1},\bm{\phi}_{j}\right).\label{eq:total_path_probability}
\end{align}

Here, $\mathcal{G}\left(\bm{\phi}_{j-1},\bm{\phi}_{j}\right)=\mathcal{G}_{j-1,j}$
is the probability of a one-step transition from $\bm{\phi}_{j-1,j}\rightarrow\bm{\phi}_{j,j-1},$
over the time interval $\epsilon$, which is:

\begin{align}
\mathcal{G}_{j-1,j} & =\int\frac{N_{j}d\bm{k}d\bm{\Delta}_{j}}{\left(\epsilon d\right)^{M}\left(2\pi\right)^{3M}}e^{-i\bm{k}\cdot\left(\bm{v}_{j}\epsilon-\bm{\Delta}_{j}\right)-\left|\bm{\Delta}_{j}\right|^{2}/\left(2\epsilon d\right)}.\label{eq:Fourier-transform-path-w}
\end{align}
This is integrated over $\bm{\Delta}_{j}$ by completing the square
to give:
\begin{align}
\mathcal{G}_{j-1,j} & =\int\frac{N_{j}d\bm{k}}{\left(2\pi\right)^{2M}}e^{-\epsilon\left[d\left|\bm{k}\right|^{2}/2+i\bm{k}\cdot\bm{v}_{j}\right]}\label{eq:Fourier-transform-path}
\end{align}

Integrating over $\bm{k}$ gives an action principle for the path
probability of the time-symmetric stochastic equation, in the form:
\begin{align}
\mathcal{G}\left[\bm{\phi}\left|\bm{\phi}_{0n}\right.\right] & =e^{-S_{0n}}.
\end{align}
{The time-symmetric }action from $t=t_{j}$ to $t=t_{n}$
is defined generally as 
\begin{equation}
{S_{jn}=\sum_{k=j+1}^{n}S_{k-1,k},}
\end{equation}
where the one-step action for the $j$-th step is $S_{j-1,j}=S_{j-1,j}^{x}+S_{j-1,j}^{y}$,
with 
\begin{align}
S_{j-1,j}^{x,y} & =S^{x,y}\left(\bm{\phi}_{j-1},\bm{\phi}_{j}\right)\nonumber \\
 & =\frac{\epsilon}{2d}\left|\bm{v}_{j}^{x,y}\right|^{2}+\ln\left(\sqrt{\mathcal{N}}/N_{j}^{x,y}\right),
\end{align}
 The normalization factor is $\mathcal{N}=\left(2\pi\epsilon d\right)^{M}$.
Both type I or type II discretizations have $N_{j}^{x,y}=1$, and
reduce to the same continuous TSSDE. 

The action is modified for type III discretization. This is a second
order form where each drift term $\bar{\bm{a}}$ is evaluated at its
midpoint value of $\bm{\bar{\phi}}_{k}=\left(\bm{\phi}_{k}+\bm{\phi}_{k-1}\right)/2$.
Since this is an implicit discretization, the normalization of the
delta-function \ref{eq:Delta-function-path} as a function of the
output variables $\phi_{k,k-1}$ includes derivatives of the drift,
$\bar{\bm{a}}$. Expanding the normalization factor $N_{j}$ in an
exponential form, and including first-order terms in $\epsilon$,
 leads to a correction to the action \citep{haken1976generalized}:
\begin{align}
\bar{S}^{x,y}\left(\bm{\phi}_{k-1},\bm{\phi}_{k}\right) & =\frac{\epsilon}{2}\left[\frac{1}{d}\left|\bar{\bm{v}}_{k}^{x,y}\right|^{2}+\nabla_{x,y}\bar{\bm{a}}_{k}^{x,y}\right]+\frac{\ln\mathcal{N}}{2},
\end{align}
where: 
\begin{align}
\bar{\bm{v}}_{k}^{x} & \equiv\frac{\bm{x}_{k}-\bm{x}_{k-1}}{\epsilon}-\bm{a}^{x}\left(\bar{\bm{\phi}}_{k}\right)\nonumber \\
\bar{\bm{v}}_{k}^{y} & \equiv\frac{\bm{y}_{k-1}-\bm{y}_{k}}{\epsilon}-\bm{a}^{y}\left(\bar{\bm{\phi}}_{k}\right).
\end{align}

These results are for the constant diffusion case, and need modification
if the diffusion depends on $\bm{\phi}.$

\section{Q-function and path-integral\label{sec:Q-function-and-path-integral}}

Next, we show that the integrated transition probability defined as
a path integral gives a time-symmetric propagator, which therefore
can generate a Q-function solution. This requires showing that the
path integral form proposed for the TSP does satisfy the Q-function
differential equation.

\subsection{TSP equivalence theorem: }

The TSP, $G\left(\bm{\phi},t_{j}\left|\bm{\phi}_{0n},\bm{t}\right.\right)$,
is obtained by integrating the $n$-step path probability $\mathcal{G}_{n}\left[\bm{\phi}\right]$
over all fields except $\bm{\phi}{}_{j}$ and the inputs $\bm{\phi}_{0n}$.
Explicitly, 
\begin{equation}
G\left(\bm{\phi},t_{j}\left|\bm{\phi}_{0n},\bm{t}\right.\right)=\int\mathcal{D}_{0n}\left[\bm{\phi}\right]\mathcal{G}\left[\bm{\phi}\left|\bm{\phi}_{0n}\right.\right]\delta\left(\bm{\phi}-\bm{\phi}_{j}\right),\label{eq:TSP-from-path integral}
\end{equation}
where the lower indices on the path integral measure define the fixed
endpoints:
\begin{equation}
\mathcal{D}_{0n}\left[\bm{\phi}\right]\equiv d\bm{y}_{0}d\bm{x}_{n}\prod_{k=1}^{n-1}d\bm{\phi}_{k}\,.
\end{equation}

\subsubsection*{Corollary: }

Provided the Q-function solution exists in integral form at initial
time $t_{0}$, the solution at all times is an integrated path probability,
multiplied by the joint probability of the input fields $\bm{\phi}_{0n}$:
\begin{equation}
Q\left(\bm{\phi},t_{j}\right)=\int\mathcal{D}\left[\bm{\phi}\right]\delta\left(\bm{\phi}-\bm{\phi}_{j}\right)\mathcal{G}_{n}\left[\bm{\phi}\right]P\left(\bm{\phi}_{0n}\right).\label{eq:symmetric-path-integral}
\end{equation}
where the path integral measure is over all the coordinates
\begin{equation}
\mathcal{D}\left[\bm{\phi}\right]\equiv\prod_{k=0}^{n}d\bm{\phi}_{k}.
\end{equation}

\subsubsection{Proof:}

To demonstrate that integrating the conditional path probability $\mathcal{G}_{n}\left[\bm{\phi}\right]$
gives a time-symmetric propagator, requires showing that the propagator
defined in (\ref{eq:TSP-from-path integral}) satisfies the time-evolution
equation, (\ref{eq:TSP-equation}), and has marginal delta-function
boundaries in the past and future, (\ref{eq:marginal-delta-TSP}). 

\paragraph*{General strategy:}

This a generalization of earlier proofs of the equivalence between
a path integral and a Fokker-Planck equation \citep{haken1976generalized,graham1977path}.
Defining $G_{j}\left(\bm{\phi}\right)\equiv G\left(\bm{\phi},t_{j}\left|\bm{\phi}_{0n},\bm{t}\right.\right)$,
and noting that $\mathcal{L}=\mathcal{L}^{x}-\mathcal{L}^{y}$, 
\[
G_{j+1}\left(\bm{\phi}\right)=\exp\left(\epsilon\mathcal{L}^{x}-\epsilon\mathcal{L}^{y}\right)G_{j}\left(\bm{\phi}\right)+O\left(\epsilon^{2}\right)\,.
\]

It follows from the Baker-Haussdorf theorem that after discretization
and taking the limit of small time-step $\epsilon$, the discrete
form of the differential equation (\ref{eq:TSP-evolution-equation})
can be written for $0\le j<n$ as:
\begin{equation}
\left(1+\epsilon\mathcal{L}^{y}\right)G_{j+1}\left(\bm{\phi}\right)=\left(1+\epsilon\mathcal{L}^{x}\right)G_{j}\left(\bm{\phi}\right)+O\left(\epsilon^{2}\right)\,.\label{eq:discrete}
\end{equation}
We will show that the postulated expansion in \ref{eq:TSP-from-path integral}
for the TSP at time $t=t_{j}$ satisfies (\ref{eq:discrete}) to first
order in $\delta$.  

The path-integral expansions for the TSP at times $t=t_{j}$ and $t=t_{j+1}$
are 
\begin{align}
G_{j}\left(\bm{\phi}\right) & =\int\mathcal{D}_{0n}\left[\bm{\phi}\right]e^{-S_{0n}}\delta\left(\bm{\phi}-\bm{\phi}_{j}\right)\,,\nonumber \\
G_{j+1}\left(\bm{\phi}\right) & =\int\mathcal{D}_{0n}\left[\bm{\phi}\right]e^{-S_{0n}}\delta\left(\bm{\phi}-\bm{\phi}_{j+1}\right)\,.
\end{align}

To prove (\ref{eq:discrete}), we introduce a hybrid probability for
$\bm{\phi}_{j+1,j}\equiv\left(\bm{x}_{j+1},\bm{y}_{j}\right)$, defined
as:

\begin{align}
\bar{G}\left(\bm{\phi}\right) & =\int\mathcal{D}_{on}\left[\bm{\phi}\right]e^{-S_{0n}}\delta\left(\bm{\phi}-\bm{\phi}_{j+1,j}\right).\label{eq:hybrid-probability}
\end{align}
and we will show in the following that in the limit of $\epsilon\rightarrow0$,

\begin{align}
\left(1+\epsilon\mathcal{L}^{y}\right)G_{j+1}\left(\bm{\phi}\right)=\left(1+\epsilon\mathcal{L}^{x}\right)G_{j}\left(\bm{\phi}\right) & =\bar{G}\left(\bm{\phi}\right).
\end{align}

\paragraph*{Boundary values:}

First, we must prove that~, 
\begin{align}
G\left(\bm{\phi},t_{0}\left|\bm{\phi}_{0n},\bm{t}\right.\right) & \propto\delta\left(\bm{x}-\bm{x}_{0}\right)\nonumber \\
G\left(\bm{\phi},t_{n}\left|\bm{\phi}_{0n},\bm{t}\right.\right) & \propto\delta\left(\bm{y}-\bm{y}_{n}\right).
\end{align}
However, the existence of delta-function boundary values is immediate.
From (\ref{eq:TSP-from-path integral}), since $D_{0n}\left[\bm{\phi}\right]$
has no measure including $\bm{x}_{0}$, it follows immediately that:
\begin{align}
G\left(\bm{\phi},t_{0}\left|\bm{\phi}_{0n},\bm{t}\right.\right) & =\int\mathcal{D}_{0n}\left[\bm{\phi}\right]\mathcal{G}_{n}\left[\bm{\phi}\right]\delta\left(\bm{\phi}-\bm{\phi}_{0}\right)\\
 & =\delta\left(\bm{x}-\bm{x}_{0}\right)\int\mathcal{D}_{0n}\left[\bm{\phi}\right]\mathcal{G}_{n}\left[\bm{\phi}\right]\delta\left(\bm{y}-\bm{y}_{0}\right).\nonumber 
\end{align}
Similarly, since $D_{0n}\left[\bm{\phi}\right]$ has no measure including
$\bm{y}_{n}$, 
\[
G\left(\bm{\phi},t_{n}\left|\bm{\phi}_{0n},\bm{t}\right.\right)=\delta\left(\bm{y}-\bm{y}_{n}\right)\int\mathcal{D}_{0n}\left[\bm{\phi}\right]\mathcal{G}_{n}\left[\bm{\phi}\right]\delta\left(\bm{x}-\bm{x}_{n}\right).
\]

\paragraph*{Single-step probability identity:}

We next prove a differential identity for the single-step transition
probability in $\bm{x}$, which factorizes from the transition probability
in $\bm{y}$ at short times. As $\epsilon\rightarrow0$, using the
Fourier representation of the step probability in Eq (\ref{eq:Fourier-transform-path}),
the single step transition probability in $\bm{x}$ is:
\begin{equation}
e^{-S_{j,j+1}^{x}}=\int\frac{d\bm{k}^{x}}{\left(2\pi\right)^{M}}e^{-\epsilon\left[d\left|\bm{k}^{x}\right|^{2}/2+i\bm{k}^{x}\cdot\bm{v}_{j+1}^{x}\right]}.
\end{equation}
Recalling the definition of relative velocity in Eq (\ref{eq:velocity}),
we use a type II discretization, and define the drift for $S_{j,j+1}^{x}$
as
\begin{align}
\bm{a}_{j}^{x} & \equiv\bm{a}^{x}\left(\bm{\phi}_{j}\right)\nonumber \\
\bm{a}_{j}^{y} & \equiv\bm{a}^{y}\left(\bm{\phi}_{j+1}\right).
\end{align}
As a result, this can be written as: 

\begin{align}
e^{-S_{j,j+1}^{x}} & =\int\frac{d\bm{k}^{x}}{\left(2\pi\right)^{M}}e^{-\epsilon\left(\frac{d}{2}\left|\bm{k}^{x}\right|^{2}-i\bm{k}^{x}\cdot\bm{a}_{j}^{x}\right)-i\bm{k}^{x}\cdot\left(\bm{x}_{j+1}-\bm{x}_{j}\right)}.
\end{align}

Expanding to first order in $\epsilon,$ and taking derivatives of
the exponential with respect to $\bm{x}_{j}$, gives the identity
that: 

\begin{align}
e^{-S_{j,j+1}^{x}} & =\left[1+\epsilon\tilde{\mathcal{L}}_{j}^{x}\right]\int\frac{d\bm{k}^{x}}{\left(2\pi\right)^{M}}e^{-i\bm{k}^{x}\cdot\left(\bm{x}_{j+1}-\bm{x}_{j}\right)}.
\end{align}
where $\mathcal{\tilde{L}}_{j}^{x}$ is the adjoint of the differential
operator $\mathcal{L}^{x}$ evaluated at $x_{j}$:
\begin{equation}
\tilde{\mathcal{L}}_{j}^{x}\equiv\frac{d}{2}\nabla_{x_{j}}^{2}+\bm{a}_{j}^{x}\cdot\bm{\nabla}_{x_{j}}.
\end{equation}

Equivalently, on carrying out the inverse Fourier transform,

\[
e^{-S_{j,j+1}^{x}}=\left[1+\epsilon\tilde{\mathcal{L}}_{j}^{x}\right]\delta\left(\bm{x}_{j}-\bm{x}_{j+1}\right)+O\left(\epsilon^{2}\right).
\]

\paragraph*{Hybrid probability:}

Inserting this expression into the expansion of the hybrid probability,
(\ref{eq:hybrid-probability}), and integrating twice by parts (assuming
vanishing probability at the phase-space boundaries), leads to
\begin{align}
\bar{G}\left(\bm{\phi}\right) & =\int\mathcal{D}_{on}\left[\bm{\phi}\right]\delta\left(\bm{x}_{j}-\bm{x}_{j+1}\right)\delta\left(\bm{\phi}-\bm{\phi}_{j+1,j}\right)\nonumber \\
 & \,\,\,\times\left[1+\epsilon\mathcal{L}_{j}^{x}\right]e^{-S_{0n}^{y}-S_{0j}^{x}-S_{j+1,n}^{x}}.
\end{align}
where:
\begin{equation}
\mathcal{L}_{j}^{x}\equiv\frac{d}{2}\nabla_{x_{j}}^{2}-\bm{\nabla}_{x_{j}}\cdot\bm{a}_{j}^{x}.
\end{equation}

On integration over $\bm{x}_{j}$, using the delta-function in $\bm{x}_{j}$,
all terms in the path integral involving $\bm{x}_{j}$ can be replaced
by $\bm{x}_{j+1}$. The $\bm{x}$ coordinates are now relabeled so
that $\bm{x}'_{k}=\bm{x}{}_{k}$ for $k<j$ and $\bm{x}'_{k-1}=\bm{x}{}_{k}$
for $k>j$ . As a result of the second delta-function in $\bm{x}$,
inside the integral $\bm{x}_{j+1}\equiv\bm{x}_{j}^{\prime}=\bm{x}$,
so that
\begin{align}
\bar{G}\left(\bm{\phi}\right) & =\left[1+\epsilon\mathcal{L}^{x}\right]\int\mathcal{N}^{n}\left(\prod_{k=1}^{n-1}d\bm{x}_{k}^{\prime}\right)\left(\prod_{k=0}^{n-1}d\bm{y}_{k}\right)\,\nonumber \\
 & \,\,\,\times\delta\left(\bm{\phi}-\bm{\phi}_{j}^{\prime}\right)e^{-S_{0,n}^{\prime y}-S_{0,n-1}^{\prime x}}.
\end{align}

This has a contracted time path in $\bm{x}$, since one of the $\bm{x}$
integrals and its corresponding propagator is now removed.

\paragraph*{Hybrid discretization:}

Here, $\bm{\phi}_{k}^{\prime}\equiv\left(\bm{x}_{k}^{\prime},\bm{y}_{k}\right),$
and a corresponding action for $S_{k,k+1}^{\prime x}$ is defined
with a drift in $x$, $y$ that is as previously for $k<j$. For $k\ge j$
one now has a type I discretization, since:
\begin{align}
\bm{a}_{k}^{\prime x} & =\bm{a}_{k+1}^{x}=\bm{a}^{x}\left(\bm{x}_{k}^{\prime},\bm{y}_{k+1}\right)\nonumber \\
\bm{a}_{k}^{\prime y} & =\bm{a}_{k}^{y}=\bm{a}^{x}\left(\bm{x}_{k}^{\prime},\bm{y}_{k+1}\right).
\end{align}
This means that the hybrid propagator has a hybrid discretization,
part type I and part type II, with 
\begin{equation}
S_{k-1,k}^{\prime}=\frac{\epsilon}{2}\left[\bm{v}_{k}^{\prime T}\left(\bm{d}^{x}\right)^{-1}\bm{v}_{k}^{\prime}\right]+\ln\mathcal{N}.
\end{equation}

To keep the total number of integration variables the same, a new
output variable $\bm{x}'_{n}$ at $t=t_{n}$ is added to the integration
measure, with a corresponding action. The new drift terms are independent
of $\bm{x}'_{n}$, as the modified action is a type I action, since
$n\ge j$. 

The corresponding integration weighted by $S_{n-1,n}^{x\prime}$ is
normalized, and equals unity. Hence, adding the new variable and integration
leaves $\bar{G}\left(\bm{\phi}\right)$ invariant. Since type I and
type II discretizations are identical in the limit of $\epsilon\rightarrow0$,
we therefore have proved that:
\begin{equation}
\bar{G}\left(\bm{\phi}\right)=\left[1+\epsilon\mathcal{L}^{x}\right]G_{j}\left(\bm{\phi}\right)\,.
\end{equation}
 Next, carrying out the complementary procedure for the $\bm{y}$
variable, we obtain a contraction in the number of $\bm{y}$ variables,
with a new variable added called $\bm{y}_{0}^{\prime}$. This also
has a type I action, and integrates to unity. The calculation is a
mirror image of the one above, leading to the required result:
\begin{equation}
\bar{G}\left(\bm{\phi}\right)=\left[1+\epsilon\mathcal{L}^{y}\right]G_{j+1}\left(\bm{\phi}\right)\,.
\end{equation}

This proves the required differential identity.

\subsection{Time-symmetric quantum action principle}

The results above show that the probability density for quantum time-evolution
is given by a path integral over a real Lagrangian, where in each
small time interval the propagators factorize. These equations can
be solved using path integrals over both the propagators.

The path then no longer has to be over an infinitesimal distance in
time, and the \emph{total} propagators will not factorize. This is
a type of stochastic bridge \citep{schrodinger1931uber,hairer2007analysis,drummond2017forward},
which acts in two time directions simultaneously. The action functional
$S$ is an integral over an effective Lagrangian, which is equivalent
to a discrete sum in the limit of small time-steps: 
\begin{equation}
\lim_{\epsilon\rightarrow0}S\left[\bm{\phi}\right]=\int_{t_{0}}^{t_{f}}L\left(\bm{\phi},\dot{\bm{\phi}}\right)dt+n\ln\mathcal{N},
\end{equation}
with a number of steps $n$ inverse to the step-size, so that $n\equiv\left(t_{f}-t_{0}\right)/\epsilon$. We
note that the stochastic equation probabilities are independent of
the type of discretization, so any discretization is possible. This
continuum limit is most readily obtained for the  symmetric type III
discretization, which is known to reach a continuum limit uniformly
for forward-time path integrals \citep{dekker1978proof,Graham1977Covariant,graham1977lagrangian},
allowing the use of standard calculus.

In order to write the action in a unified form, we define a combined,
central difference Lagrangian as
\begin{equation}
L=L^{x}(\bm{\phi},\dot{\bm{\phi}})+L^{y}(\bm{\phi},-\dot{\bm{\phi}})\,,
\end{equation}
so that the action integral can be written in the positive time direction
for $t_{0}<t<t_{f}$, with a total Lagrangian of
\begin{align}
L & =\sum_{\mu}\frac{1}{2d}\left(\dot{\phi}^{\mu}-A^{\mu}\left(\bm{\phi},t\right)\right)^{2}-V\left(\bm{\phi}\right).\label{eq:Total Lagrangian}
\end{align}
Here the potential term $V$ includes contributions of opposite sign
from the positive and negative time fields, so that
\begin{equation}
V\left(\bm{\phi},t\right)=-\frac{1}{2}\sum_{\mu}\partial_{\mu}a^{\mu}\left(\bm{\phi}\right).\label{eq:Jacobian correction term-2}
\end{equation}

This defines the total probability for an $n$-step open stochastic
bidirectional bridge, with constant diffusion, central difference
evaluation of the action, and fixed intermediate points:
\begin{equation}
G\left(\left[\bm{\phi}\right]\left|\bm{\phi}_{IN}\right.\right)=\mathcal{N}^{n}e^{-\int_{t_{0}}^{t_{f}}L(\bm{\phi},\dot{\bm{\phi}})dt}.\label{eq:probability_solution}
\end{equation}

On integrating over the intermediate points, with drift terms defined
at the center of each step in phase-space, this can be written in
a notation analogous to a quantum-mechanical transition amplitude
in a Feynman path integral. One obtains the Q-function in this limit
as: 

\[
Q\left(\bm{\phi}',t'\right)=\int d\mu\left[\bm{\phi}\right]\delta\left(\bm{\phi}\left(t'\right)-\bm{\phi}'\right)e^{-\int_{t_{0}}^{t_{f}}L(\bm{\phi},\dot{\bm{\phi}})dt'}P\left(\bm{\phi}_{IN}\right)
\]
where $t_{n}=t_{0}+n\epsilon$, and we integrate over all phase-space
points with a normalized measure:
\begin{equation}
d\mu\left[\bm{\phi}\right]=\lim_{\epsilon\rightarrow0}\mathcal{N}^{n}\prod_{k=0}^{n}d\bm{\phi}_{k}\,.
\end{equation}
The paths $\bm{\phi}\left(t\right)$ are defined so that $\bm{\phi}_{IN}=\left(\bm{x}_{0},\bm{y}_{f}\right)$,
where $\bm{x}\left(t_{0}\right)=\bm{x}_{0}$ and $\bm{y}\left(t_{n}\right)=\bm{y}_{f}$
are defined at the initial and final times respectively. 

\section{Extra dimensions\label{sec:Extra-dimensions}}

Many techniques exist for evaluating real path-integrals, both numerical
and analytic. There is a formal analogy between the form given above
and the expression for a Euclidean path integral of a polymer, or
a charged particle in a magnetic field. Here we obtain an extra-dimensional
technique for probabilistic sampling of the time-symmetric path integral,
which also gives an algorithm for evaluating the solution to a TSSDE.
Similar results are known for forward time stochastic equations \citep{hairer2007analysis,drummond2017forward}.

\subsection{Equilibration in higher dimensions}

To make use of the real path-integral, one needs to probabilistically
sample the entire space-time path, since each part of the path depends
in general on other parts. To achieve this, we add an additional 'virtual'
time dimension, $\tau$. This is used in the related statistical problem
of stochastic bridges, for computing a stochastic trajectory that
is constrained by a future boundary condition \citep{schrodinger1931uber,maier1992transition,hairer2007analysis,majumdar2015effective,drummond2017forward}.

This extra-dimensional functional distribution, $\mathcal{P}\left([\bm{\phi}],\tau\right)$,
is defined so that the probability tends asymptotically for large
$\tau$ to the required solution:
\begin{equation}
\lim_{\tau\rightarrow\infty}\mathcal{P}\left([\bm{\phi}],\tau\right)=\mathcal{G}\left(\left[\bm{\phi}\right]\left|\bm{\phi}_{IN}\right.\right)\,.
\end{equation}
The solution is such that $\bm{\phi}\left(t\right)$ is constrained
so that $\bm{x}\left(t_{0}\right)=\bm{x}_{0}$ , and $\bm{y}\left(t_{f}\right)=\bm{y}_{f}$,
where $\bm{x}_{0},\bm{y}_{f}$ are randomly distributed as $\mathcal{P}\left(\bm{x}_{0},\bm{y}_{f}\right)$
for the case of an nonlocal input boundary. We  use the Type III midpoint
form of the Lagrangian.

It has been shown in work on stochastic bridges \citep{hairer2007analysis}
that sampling using a stochastic partial differential equation (SPDE)
can be applied to cases where one of the boundary conditions is free.
To define an SPDE the other boundary condition on $\bm{x}$ is specified
so that $\dot{\bm{x}}\left(t_{f}\right)=\bm{a}^{x}\left(\bm{\phi}\left(t_{f}\right)\right)$,
with a boundary condition for $\dot{\bm{y}}$ so that $\dot{\bm{y}}\left(t_{0}\right)=-\bm{a}^{y}\left(\bm{\phi}\left(t_{0}\right)\right)$.

This is consistent with the open boundary conditions of the path integral
in real time, since in the limit of $\epsilon\rightarrow0$, the path
integral weight implies that one must have $\dot{\bm{x}}\left(t_{f}\right)=\bm{a}^{x}\left(\bm{\phi}\left(t_{f}\right)\right)+O\left(\sqrt{\epsilon}\right)$
and $\dot{\bm{y}}\left(t_{0}\right)=-\bm{a}^{y}\left(\bm{\phi}\left(t_{0}\right)\right)+O\left(\sqrt{\epsilon}\right)$.
The effect of the additional terms proportional to $\sqrt{\epsilon}$
vanish as $\epsilon\rightarrow0$, as they contributes a negligible
change to the entire path integral. This open boundary condition is
necessary in order to have a well-defined partial differential equation
in higher dimensions.

Extra-dimensional equilibration is seldom used for conventional SDE
sampling, as direct evolution is more efficient. However, we will
show that SPDE sampling is applicable to time-symmetric propagation,
where direct sampling is not possible without additional iteration.
In this section, a simplification is made by rescaling the variables
to make the diffusion $d^{\mu}\left(t\right)$ independent of time
and index, i.e., $d^{\mu}\left(t\right)=d$, as in the previous section. 

The SPDE is obtained as follows \citep{drummond2017forward}. Firstly
suppose that $\mathcal{P}\left([\bm{\phi}],\tau\right)$ satisfies
a functional partial differential equation of
\begin{equation}
\frac{\partial\mathcal{P}}{\partial\tau}=\int_{t_{0}}^{t_{f}}dt\sum_{\mu}\frac{\delta}{\delta\phi^{\mu}(t)}\left[-\mathcal{A}^{\mu}\left(\bm{\phi},t\right)+d\frac{\delta}{\delta\phi^{\mu}(t)}\right]\mathcal{P}\,.\label{eq:functional FPE}
\end{equation}
In order that the asymptotic result agrees with the desired expression
for $\mathcal{G}$, it follows from functional differentiation of
Eq (\ref{eq:total_path_probability}), that one must define $\bm{\mathcal{A}}\left(\bm{\phi},t\right)$
so that
\begin{equation}
\mathcal{A}^{\mu}\left(\bm{\phi},t\right)=-d\frac{\delta}{\delta\phi^{\mu}(t)}\int_{t_{0}}^{t_{f}}L(\bm{\phi},\dot{\bm{\phi}})dt'.
\end{equation}

This is a variational calculus problem, with one boundary fixed, and
the other free. Variations vanish at the time boundaries where $\bm{\phi}$
is fixed. At the free boundaries, we choose that $\dot{\bm{\phi}}=\bm{A}$.
In either case, boundary terms are zero because they occur in terms
that vanish provided
\begin{equation}
\Delta\phi^{\nu}\frac{\partial L}{\partial\dot{\phi}^{\mu}}=\frac{\Delta\phi^{\nu}}{d}\left(\dot{\phi}^{\mu}-A^{\mu}\right)=0\,.
\end{equation}
As a result, there are two type of natural boundary terms that allow
partial integration to obtain Euler-Lagrange equations. Either one
can set $\Delta\phi^{\mu}=0$ to give a fixed Dirichlet boundary term,
or else one can set $\dot{\phi}^{\mu}=A^{\mu}$, to give an open Neumann
boundary term. Hence, we choose to set $x^{j}=x_{0}^{j}$ and $\dot{y}^{j}=A^{yj}$
at $t=t_{0}$, while $y^{j}=y_{f}^{j}$ and $\dot{x}^{j}=A^{xj}$
at $t=t_{f}$. This allows one to obtain Euler-Lagrange type equations
with an extra-dimensional drift defined as
\begin{align}
\mathcal{A}^{\mu}\left(\bm{\phi},t\right) & =d\left[\frac{d}{dt}\frac{\partial L}{\partial\dot{\phi}^{\mu}}-\frac{\partial L}{\partial\phi^{\mu}}\right]\nonumber \\
 & =\frac{d}{dt}\left(\dot{\phi}^{\mu}-A^{\mu}\right)+\left(\dot{\phi}^{\nu}-A^{\nu}\right)\partial_{\mu}A^{\nu}+d\partial_{\mu}V\,.
\end{align}

The functional Fokker-Planck equation given above is then equivalent
to a stochastic partial differential equation (SPDE):
\begin{equation}
\frac{\partial\bm{\phi}}{\partial\tau}=\bm{\mathcal{A}}\left(\bm{\phi},t\right)+\bm{\zeta}\left(t,\tau\right)\,,\label{eq:SPDE}
\end{equation}
where the stochastic term $\bm{\zeta}$ is a real delta-correlated
Gaussian noise such that 
\begin{equation}
\left\langle \zeta^{\mu}\left(t,\tau\right)\zeta^{\mu}\left(t',\tau'\right)\right\rangle =2d\delta^{\mu\nu}\delta\left(t-t'\right)\delta\left(\tau-\tau'\right).
\end{equation}

\subsection{Coefficients}

Introducing first and second derivatives, $\dot{\bm{\phi}}\equiv\partial\bm{\phi}/\partial t$
and $\ddot{\bm{\phi}}\equiv\partial^{2}\bm{\phi}/\partial t^{2}$,
there is an expansion for the higher-dimensional drift term $\bm{\mathcal{A}}$
in terms of the field time-derivatives:
\begin{equation}
\bm{\mathcal{A}}\left(\bm{\phi},t\right)=\ddot{\bm{\phi}}+\bm{C}\dot{\bm{\phi}}+\bm{U}\,.
\end{equation}
Here, $\bm{C}$ is a circulation matrix that only exists when the
usual potential conditions on the drift are not satisfied \citep{drummond2017forward},
while $\bm{U}$ is a pure drift without time-derivatives:
\begin{align}
C^{\mu\nu} & =\partial_{\mu}A^{\nu}-\partial_{\nu}A^{\mu}\,.\\
U^{\mu} & =\partial_{\mu}\left(dV-\frac{1}{2}\sum_{\nu}\left(A^{\nu}\right)^{2}\right)\,.\nonumber 
\end{align}

The function $U$ acts to generate an effective force on the trajectories.
The final stochastic partial differential equation that $\bm{\phi}$
must satisfy is therefore
\begin{equation}
\frac{\partial\bm{\phi}}{\partial\tau}=\ddot{\bm{\phi}}+\bm{C}\dot{\bm{\phi}}+\bm{U}+\bm{\zeta}\left(t,\tau\right)\,.\label{eq:finalPSDE}
\end{equation}

The final result is a c-number field stochastic partial differential
equation in an extra space-time dimension, including an additional
noise term. It has a steady-state that is equivalent to a full quantum
evolution equation, and is identical to classical evolution in real
time in the zero-noise limit, as shown in the next subsection. 

The equations can be treated with standard techniques for stochastic
partial differential equations \citep{Werner:1997}. The equations
have $n_{d}+1$ dimensions in a manifold with $n_{d}$ space-time
dimensions. The simplest case, for a single mode, has $n_{d}+1=2$
dimensions. In computational implementations, one can speed up convergence
to the steady-state using Monte-Carlo acceleration \citep{besag1994comments}.

\subsection{Classical~limit}

The classical limit is for $d\rightarrow0$. In this limit the higher-dimensional
equations are noise-free and diffusive. Ignoring the noise term, one
obtains:
\begin{equation}
\frac{\partial\phi^{\mu}}{\partial\tau}=\ddot{\phi}^{\mu}+\left[\partial_{\mu}A^{\nu}-\partial_{\nu}A^{\mu}\right]\dot{\phi}^{\nu}-A^{\nu}\partial_{\mu}A^{\nu}\,.
\end{equation}
We wish to show that the classical trajectory solution, $\dot{\phi}^{\nu}=A^{\nu}$,
is a possible steady-state solution. In this case, we obtain for these
trajectories,
\begin{equation}
\frac{\partial\phi^{\mu}}{\partial\tau}=\frac{d}{dt}A^{\mu}-A^{\nu}\partial_{\nu}A^{\mu}.
\end{equation}

Therefore, on a noise-free trajectory, the second derivative term
simplifies to give
\[
\frac{\partial\phi^{\mu}}{\partial\tau}=\left[\dot{\phi}^{\nu}-A^{\nu}\right]\partial_{\nu}A^{\mu}.
\]
and hence for the classical trajectory one obtains
\begin{equation}
\frac{\partial\phi^{\mu}}{\partial\tau}=\mathcal{A}^{\mu}\left(\bm{\phi}\right)=0\,.\label{eq:classical_limit}
\end{equation}
This extra-dimensional equation therefore has a steady state solution
corresponding to the integrated classical field evolution in real
time, namely:
\begin{align}
\bm{x}(t) & =\bm{x}(t_{0})+\int_{t_{0}}^{t}\bm{a}^{x}\left(\bm{\phi}\left(t'\right)\right)dt'\nonumber \\
\bm{y}(t) & =\bm{y}(t_{f})+\int_{t}^{t_{f}}\bm{a}^{y}\left(\bm{\phi}\left(t'\right)\right)dt'\,.
\end{align}

Both the initial and final boundary term equations are satisfied provided
one chooses $\bm{x}\left(t_{0}\right)=\bm{x}_{0}$ and $\bm{y}\left(t_{f}\right)=\bm{y}_{f}$,
if these are compatible, that is, if the dynamical equations have
a solution. If one uses these equations to solve for $\bm{y}(t_{0})$,
the solution can be rewritten in a more conventional form of a classical
solution with initial conditions:
\begin{equation}
\bm{\phi}\left(t\right)=\bm{\phi}\left(t_{0}\right)+\int_{t_{0}}^{t}\bm{A}\left(\bm{\phi}\left(t'\right)\right)dt'.
\end{equation}

The importance of imposing future-time boundary conditions in classical
field problems like radiation-reaction has long been recognized in
electrodynamics, including work by Dirac \citep{dirac1938pam}, as
well as Wheeler and Feynman \citep{wheeler1945interaction}. In such
theories various field components typically require future-time restrictions
on their dynamics. Hence the fact that future-time boundaries arise
in the classical limit found here should not be very surprising. 

Dirac \citep{dirac1938pam} described his result that effectively
gives a future boundary condition on electron acceleration in as ``\emph{the
most beautiful feature of the theory}''. He explains: ``\emph{We
now have a striking departure from the usual ideas of mechanics. We
must obtain solutions of our equations of motion for which the initial
position and velocity of the electron are prescribed, together with
its final acceleration, instead of solutions with all the initial
conditions prescribed.''}

If Dirac's theory is compared with the classical limit obtained here,
there are are clear similarities. His approach gave a dynamical condition
required to derive the correct time evolution, using a restriction
on the \emph{future} boundaries of the radiation field. It is a striking
feature of the present approach that Dirac's idea of a future boundary
condition arises naturally from the zero-noise limit of our equations.

\subsection{Numerical methods}

A variety of numerical techniques can be used to implement path integrals
with a time-symmetric action. In this paper we solve the equivalent
higher-dimensional partial stochastic differential equation with a
finite difference implementation. This permits Neumann, Dirichlet
and other boundary conditions to be imposed. We also explain strategies
for dealing with future time boundaries, which is the most obvious
practical issue with this approach.

\subsubsection{SPDE integration}

First, we demonstrate convergence of the higher dimensional method,
for an ordinary stochastic differential equation. We use a central
difference implicit method that iterates to obtain convergence at
each step, including an iteration of the boundary conditions. The
method is similar to a central difference method described elsewhere
\citep{drummond1991computer,Werner:1997}. A simple finite difference
implementation of the Laplacian is used to implement non-periodic
time boundaries.

\begin{figure}
\includegraphics[width=1\columnwidth]{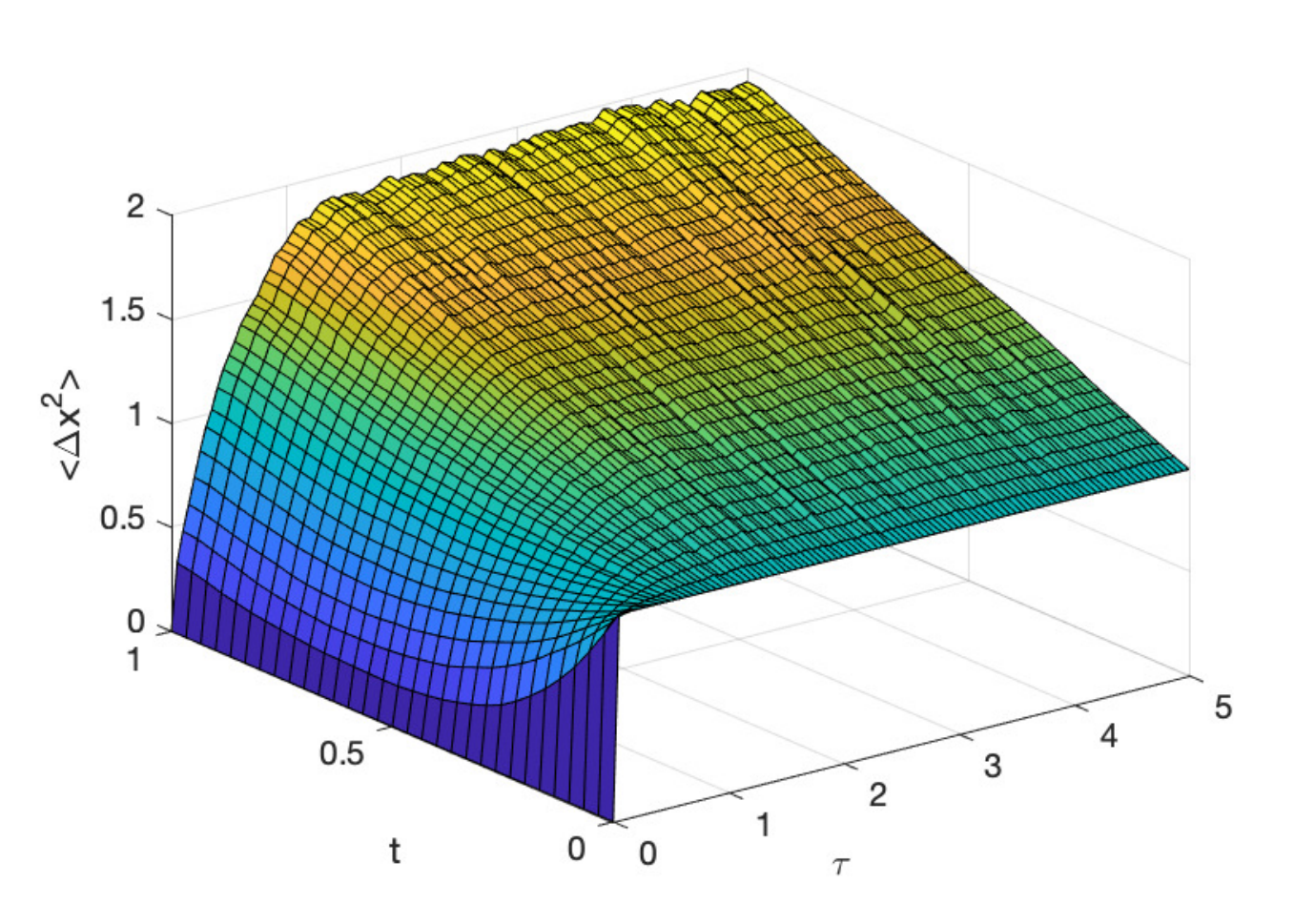}\caption{Example of SPDE solution with an extra dimension. The component $x$
propagates in the positive time direction as a random Wiener process.
The expected variance for $\tau\rightarrow\infty$ is $\left\langle x^{2}\left(t,\tau\right)\right\rangle =1+t$,
with $\left\langle x^{2}\left(0,\tau\right)\right\rangle =1$. Fluctuations
are sampling errors due to a finite number of $10000$ trajectories.
Variance error bars due to sampling errors were estimated as $\pm2.5\%$,
in good agreement with the difference between exact and simulated
variance. A semi-implicit finite difference method \citep{drummond1991computer,werner1997robust}
was used to integrate the SPDE in $\tau$, with step-sizes of $\Delta\tau=0.0002$
and $\Delta t=0.03$. Errors from the finite step-size in $\tau$
were negligible. Finite differences were also used for the derivatives
in $t$. \label{fig:Quantum-diffusion}}
\end{figure}

In order to demonstrate convergence, Fig (\ref{fig:Quantum-diffusion})
gives the computed numerical variance in an exactly soluble example
of a stochastic differential equation with no drift term. We treat
one variable with $\bm{C}=\bm{U}=0$, using a public-domain SPDE solver
\citep{kiesewetter2016xspde} with a random Gaussian initial condition
of $\left\langle x_{0}^{2}\right\rangle =1$, where:
\begin{equation}
x(t)=x_{0}+\int_{0}^{t}dw_{x}\,.
\end{equation}
This is a case of pure diffusion, so one expects the final equilibrium
solution as $\tau\rightarrow\infty$ to be $\left\langle x^{2}\left(t,\tau\right)\right\rangle =1+t$.
From Eq (\ref{eq:finalPSDE}), the corresponding higher-dimensional
stochastic process has boundary conditions of $x(t=0)=x_{0}$ and
$\dot{x}(t=t_{f})=0$, while satisfying a stochastic partial differential
equation:
\begin{equation}
\frac{\partial x}{\partial\tau}=\ddot{x}+\zeta\left(t,\tau\right)\,.
\end{equation}
From the numerical results in Fig (\ref{fig:Quantum-diffusion}),
the expected variance is reached uniformly in real time $t$ after
pseudo-time $\tau\sim$ 2.5, to an excellent approximation, reaching
$\left\langle x^{2}\right\rangle =1.95\pm0.05$ at $t=t_{f}=1$ and
$\tau=5$. 

For the cases treated here, our focus is on accuracy rather than numerical
efficiency. The purpose of the examples in this paper is to demonstrate
how this approach works in simple cases. Checks were made to quantitatively
estimate sampling error and step-size error in $\tau$. Substantial
improvements in efficiency appear possible. It may be feasible to
combine Ritz-Galerkin \citep{matthies2005galerkin}, spectral \citep{Werner:1997},
or other methods \citep{keese2003review} with boundary iteration.
The MALA technique for accelerated convergence is also applicable
\citep{besag1994comments}.

\section{\label{sec:Quadratic-Hamiltonian-Examples} Examples}

Hamiltonians in quantum field theory of the type analyzed here have
quadratic and quartic terms. In this section we consider two elementary
examples, with details in single-mode cases. We consider a Hamiltonian
of the form $\widehat{H}=\widehat{H}_{0}+\widehat{H}_{S}$. Here $\widehat{H}_{0}$
is a free field term, and $\widehat{H}_{S}$ describes quadrature
squeezing, found in Hawking radiation or parametric down-conversion.Each
of these cases will be treated separately below for simplicity, but
they can be combined if required. 

\subsection{Free-field case}

After discretizing on a momentum lattice, and using the Einstein summation
convention, the free-field Hamiltonian can be written in normally-ordered
form as
\begin{equation}
\widehat{H}=\hbar\omega_{ij}\hat{a}_{i}^{\dagger}\hat{a}_{j}\,.
\end{equation}
The corresponding Q-function equations are:
\begin{equation}
\dot{Q}^{\alpha}=-i\omega_{ij}\left[\frac{\partial}{\partial\alpha_{j}^{\ast}}\alpha_{i}^{\ast}-\frac{\partial}{\partial\alpha_{i}}\alpha_{j}\right]Q^{\alpha}\,.
\end{equation}
Hence, the coherent amplitude evolution equations are:
\begin{equation}
\frac{d\alpha_{i}}{dt}=-i\omega_{ij}\alpha_{j}.\label{linear_evolution}
\end{equation}

The simplest case is a single-mode simple harmonic oscillator Hamiltonian,
such that: $\widehat{H}=\hbar\omega\hat{a}^{\dagger}\hat{a}\,.$ This
corresponds to a characteristic equation of $\dot{\alpha}=-i\omega\alpha.$
The expectation value of the coherent amplitude in the Q-function
has the equation:
\begin{equation}
\frac{\partial}{\partial t}\left\langle \alpha\right\rangle _{Q}=-i\omega\left\langle \alpha\right\rangle _{Q},
\end{equation}
which is identical to the corresponding Heisenberg equation expectation
value. There is no diffusive behavior or noise for these terms, and
as a result the Q-function has an exactly soluble, deterministic quantum
dynamics. The evolution is noise-free, with no need to make the transformations
outlined above, since from (\ref{eq:classical_limit}), the steady-state
in extra dimensions is given by solving(\ref{linear_evolution}).
There is no difference between classical and quantum dynamics for
coherent states, as pointed out by Schr\"odinger \citep{Schrodinger_CS}.

\subsection{Squeezed state evolution}

Next, we consider quadratic interaction terms that are mapped to second-order
derivatives in the Q-function. These cause squeezed state generation
with quantum noise. They lead to a model for quantum measurement and
quantum paradoxes \citep{drummond2020retrocausal}.

Following the notation of Eq (\ref{eq:General hermitian}), the general
squeezing interaction term is $\widehat{H}_{S}=\hbar\sum_{ij=0}^{M}\left[g_{ij00}\hat{a}_{i}^{\dagger}\hat{a}_{j}^{\dagger}+g_{00ij}\hat{a}_{i}\hat{a}_{j}\right]/2$.
Such quadrature squeezing interactions are found in many areas of
physics \citep{Drummond2004_book}. They illustrate how the Q-function
equation behaves in the simplest nontrivial case where there is a
diffusion term that is not positive-definite. We will investigate
this in some detail, with numerical examples. This case illustrates
how complementary variance changes are related to complementary time
propagation directions.

Physically, these terms arise from parametric interactions, and lead
to the dynamics that cause quantum entanglement. They are widespread,
occurring in systems ranging from quantum optics to black holes, via
Hawking radiation. The simplest case, with $g_{ij00}=\delta_{ij}$,
is a single-mode quantum squeezing Hamiltonian: 
\begin{equation}
\widehat{H}=\frac{i\hbar}{2}\left[\hat{a}^{\dagger2}-h.c.\right]\,.
\end{equation}

\subsubsection{Q-function dynamics}

We can calculate directly how the Q-function evolves in time. Applying
the correspondence rules as previously, one obtains a time-symmetric
Fokker-Planck type equation, now with second-order terms. Combining
these terms into one equation gives:
\begin{align}
\frac{dQ^{\alpha}}{dt} & =-\left[\frac{\partial}{\partial\alpha}\alpha^{\ast}+\frac{1}{2}\frac{\partial^{2}}{\partial\alpha^{2}}+h.c\right]Q^{\alpha}\,.
\end{align}

Using the standard quadrature definitions of Eq (\ref{eq:real-quadratures})
one has $\alpha=q+ip$. The real phase space variables of Eq (\ref{eq:xy-quadratures})
with positive and negative diffusion are found by noting that $e^{i\eta}=i$,
so making a variable change with $i\alpha=x+iy=iq-p$, we obtain
\begin{equation}
\frac{dQ}{dt}=\left[\partial_{x}x-\partial_{y}y+\frac{1}{4}\left(\partial_{x}^{2}-\partial_{y}^{2}\right)\right]Q\,.
\end{equation}

This demonstrates the typical behavior of unitary Q-function equations.
The diffusion matrix is traceless and equally divided into positive
and negative definite parts. In this case the $x$ quadrature decays,
but has positive diffusion, while the the $y$ quadrature shows growth
and amplification, but has negative diffusion in the forward time
direction. The amplified quadrature, which corresponds to the measured
signal of a parametric amplifier, has a negative diffusion and is
constrained by a future time boundary condition.

If initially factorizable, the $Q$-function solutions can always
be factorized as a product with $Q=Q_{x}Q_{y}$. Then, if $t_{-}=t_{0}+t_{1}-t$,
the time-evolution is diffusive, with an identical structure in each
of two different time directions:
\begin{align}
\frac{dQ_{y}}{dt_{-}} & =\partial_{y}\left[-y+\frac{1}{4}\partial_{y}\right]Q_{y}\nonumber \\
\frac{dQ_{x}}{dt} & =\partial_{x}\left[-x+\frac{1}{4}\partial_{x}\right]Q_{x}\,.
\end{align}

The corresponding forward-backwards SDE is uncoupled, with decay and
stochastic noise occurring in each time direction: 
\begin{align}
x(t) & =x(t_{0})-\int_{t_{0}}^{t}x(t')dt'+\int_{t_{0}}^{t}dw_{x}\nonumber \\
y(t) & =y(t_{f})-\int_{t}^{t_{f}}y(t')dt'-\int_{t}^{t_{f}}dw_{y}\,.
\end{align}

where $\left\langle dw_{\mu}dw_{\nu}\right\rangle =\frac{1}{2}\delta_{\mu\nu}dt$.
From these equations one can calculate immediately that
\begin{align}
\frac{d}{dt}\left\langle x^{2}\right\rangle  & =2\left(\frac{1}{4}-\left\langle x^{2}\right\rangle \right)\nonumber \\
\frac{d}{dt_{-}}\left\langle y^{2}\right\rangle  & =2\left(\frac{1}{4}-\left\langle y^{2}\right\rangle \right).
\end{align}
 This equation for the variance time-evolution implies that the variance
is therefore \emph{reduced} in each quadrature's intrinsic diffusion
direction, for an initial vacuum state, with the solution in forward
time given by
\begin{align}
\left\langle x^{2}\left(t\right)\right\rangle  & =\frac{1}{4}\left(1+e^{-2t}\right)\nonumber \\
\left\langle y^{2}\left(t\right)\right\rangle  & =\frac{1}{4}\left(1+e^{2t}\right).\label{eq:Squeezed-Q-function-solution}
\end{align}
Therefore, the variance reduction occurs in the forward time direction
for $x$, giving rise to quadrature squeezing for an initial vacuum
state, and in the backward time direction for $y$, leading to gain
in the forward time direction. However, neither anti-normally ordered
variance is reduced below $1/4$. This is the minimum possible, corresponding
to zero variance in the unordered operator case.

With this choice of units, the diffusion coefficient is $d=1/2$,
so the total Lagrangian of Eq (\ref{eq:Total Lagrangian}) is:
\begin{equation}
L=\left(\dot{x}+x\right)^{2}+\left(\dot{y}-y\right)^{2}-1.
\end{equation}

\begin{figure}
\includegraphics[width=1\columnwidth]{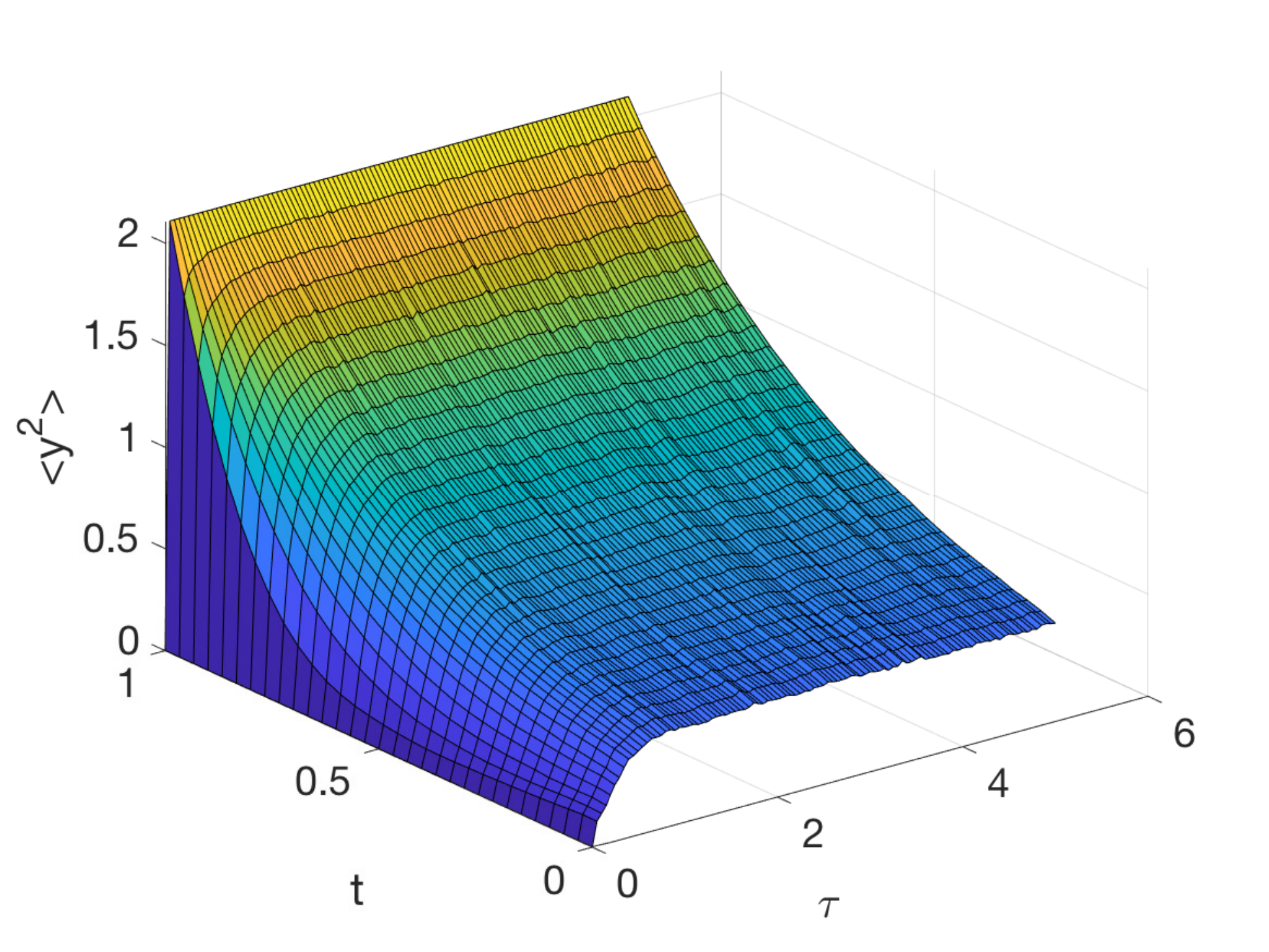}\caption{Variance of SPDE solution with an extra dimension. The unsqueezed
quadrature $y$ propagates in the negative time direction, with boundaries
fixed in the future. The extra-dimensional stochastic partial differential
equation is solved out to $\tau=5$ , with a future time Dirichlet
boundary at $t=1$ of a specified Gaussian distribution of $y$ with
the correct variance at all $\tau$, and a past time Robin boundary
at $t=0$ with a fixed derivative. Fluctuations are sampling errors
due to a finite number of $6400$ stochastic trajectories. Other details
as in Fig (\ref{fig:Quantum-diffusion}). \label{fig:Quantum-unsqueeze_vs_tau-t}}
\end{figure}

\begin{figure}
\includegraphics[width=1\columnwidth]{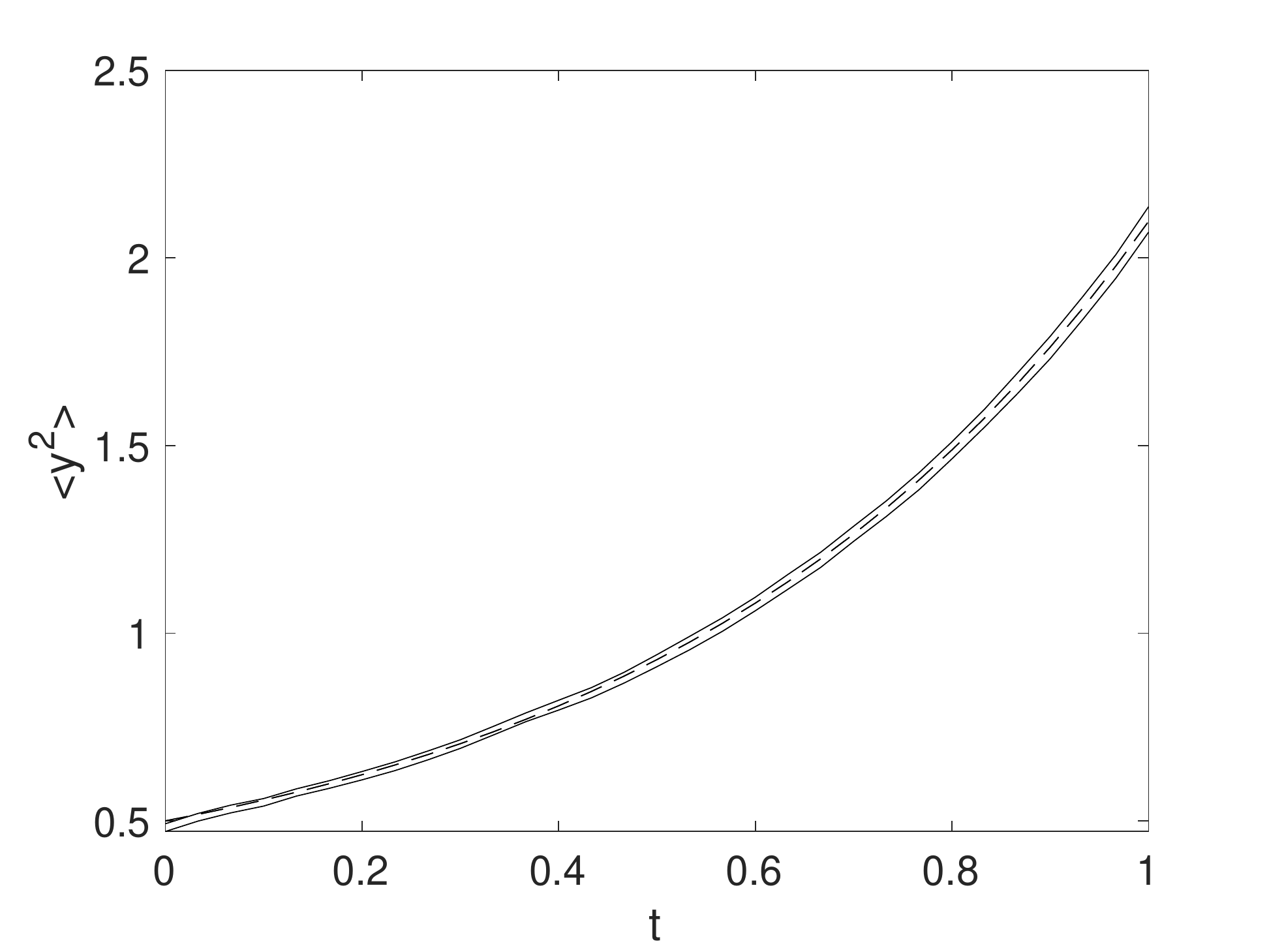}\caption{Example of SPDE solution with an extra dimension. The unsqueezed quadrature
variance $y$ propagates in the negative time direction, with results
obtained at virtual time $\tau=5$. The expected variance for $\tau\rightarrow\infty$
is $\left\langle y^{2}\left(t\right)\right\rangle =\frac{1}{4}\left(1+e^{2t}\right)$,
and is shown as the dotted line. Fluctuations are sampling errors
due to a finite number of $6400$ stochastic trajectories. The two
solid lines are plus and minus one standard deviations from the mean.
Other details as in Fig (\ref{fig:Quantum-diffusion}). \label{fig:Quantum-unsqueeze_vs_t}}
\end{figure}

\begin{figure}
\includegraphics[width=1\columnwidth]{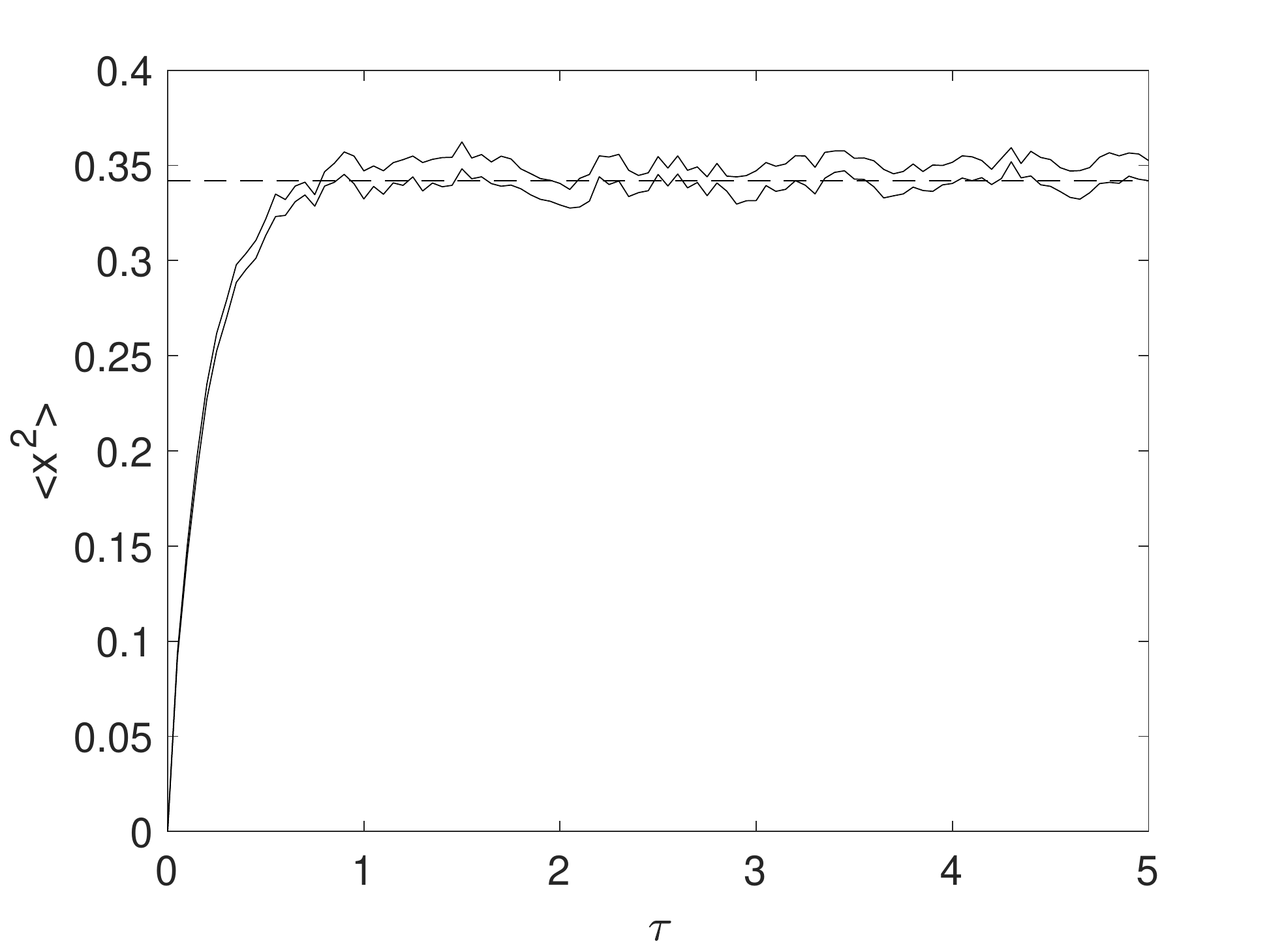}\caption{The equilibration in $\tau$ of the squeezed quadrature variance $x$
propagating in the positive time direction, with results obtained
at real time $t=0.5$. Full equilibration is achieved for $\tau\gtrsim$1.
Other details as in Fig (\ref{fig:Quantum-unsqueeze_vs_t}). \label{fig:Quantum-unsqueeze_vs_tau}}
\end{figure}

\begin{figure}
\includegraphics[width=1\columnwidth]{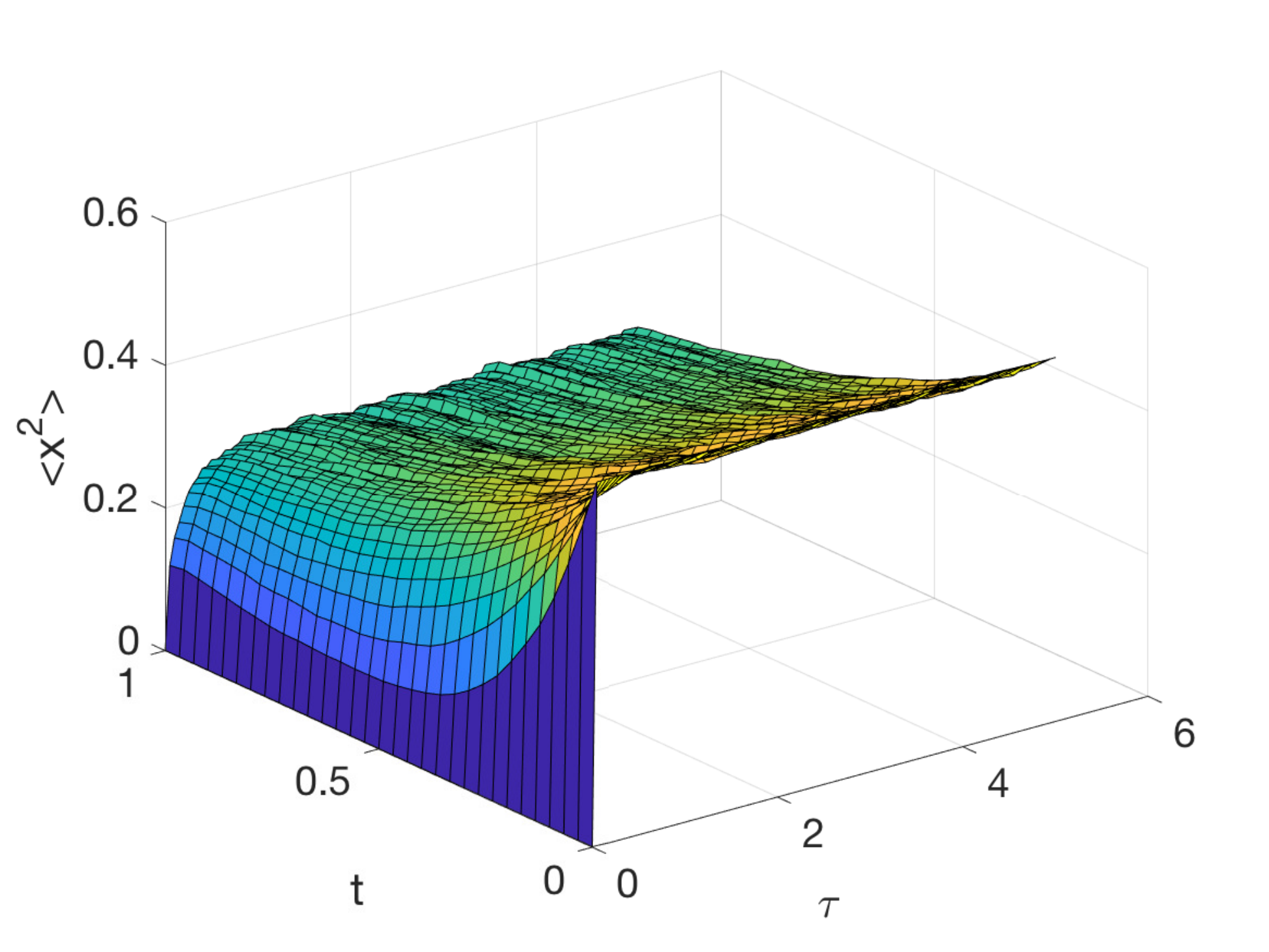}\caption{Variance of SPDE solution with an extra dimension. The squeezed quadrature
$x$ propagates in the positive time direction, using a partial stochastic
differential equation with with a past time Dirichlet boundary at
$t=0$ of a specified Gaussian distribution in $x$ with the correct
variance, and a future time Robin boundary at $t=1$ with a specified
derivative. Other details as in Fig (\ref{fig:Quantum-unsqueeze_vs_t}).
\label{fig:Quantum-squeeze_vs_tau_t}}
\end{figure}

\begin{figure}
\includegraphics[width=1\columnwidth]{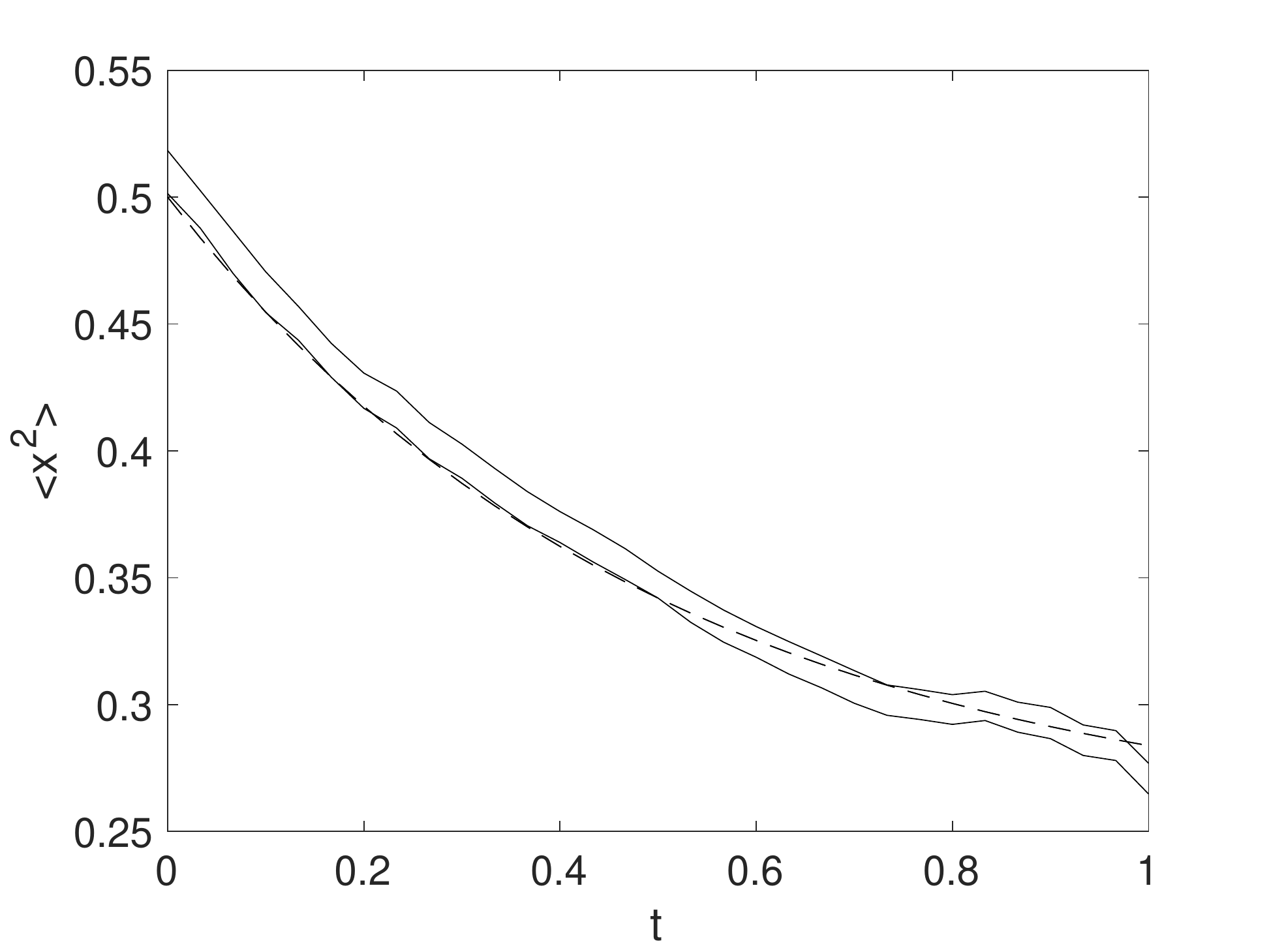}\caption{Example of SPDE solution with an extra dimension. The squeezed quadrature
variance $x$, propagates in the positive time direction, with results
obtained at virtual time $\tau=5$. The expected variance for $\tau\rightarrow\infty$
is $\left\langle x^{2}\left(t\right)\right\rangle =\frac{1}{4}\left(1+e^{-2t}\right)$,
shown as the dotted line. Fluctuations are sampling errors due to
a finite number of $1600$ stochastic trajectories. The two solid
lines are plus and minus one standard deviations from the mean. Other
details as in Fig (\ref{fig:Quantum-diffusion}). \label{fig:Quantum-squeeze_vs_t}}
\end{figure}

The net effect of the stochastic processes in opposite time directions
is that growth in the uncertainty of one quadrature in one time direction
is cancelled by the reduction in uncertainty of the other quadrature
in the opposite time direction. This behavior is shown in Figs (\ref{fig:Quantum-unsqueeze_vs_tau-t})
to (\ref{fig:Quantum-squeeze_vs_t}), which illustrate numerical solutions
of the forward-backward equations using the techniques of the previous
section. 

These solutions use 6400 trajectories, and hence include sampling
error. 

Three dimensional graphs show equilibration in the extra dimension.
Two dimensional graphs show results near equilibrium at $\tau=5$,
with plots of variance in $x,y$ vs real time $t$ and virtual time
$\tau$.

\subsection{Comparison to operator equations}

Defining quadrature operators $\hat{y}=\left(\hat{a}+\hat{a}^{\dagger}\right)/2=\hat{q}$
and $\hat{x}=i\left(\hat{a}-\hat{a}^{\dagger}\right)/2=-\hat{p}$,
this physical system has the well-known behavior that the two variances
change exponentially in time \citep{Walls1983a}, in a complementary
way. Given an initial vacuum state in which $\left\langle \hat{x}^{2}\left(0\right)\right\rangle =\left\langle \hat{y}^{2}\left(0\right)\right\rangle =1$,
the Heisenberg equation solutions for the variances are
\begin{align}
\left\langle \hat{x}^{2}\left(t\right)\right\rangle  & =\frac{1}{4}e^{-2t}\nonumber \\
\left\langle \hat{y}^{2}\left(t\right)\right\rangle  & =\frac{1}{4}e^{2t}.
\end{align}
Hence, the $\hat{x}$ quadrature is squeezed, developing a variance
below the vacuum fluctuation level, and the $\hat{y}$ quadrature
is unsqueezed, developing a large variance. This maintains the Heisenberg
uncertainty product, which is invariant.

Once operator ordering is taken into account, this gives an identical
solution to the Q-function solution in Eq \ref{eq:Squeezed-Q-function-solution},
because the operator correspondences are for anti-normal ordering.
If we use $\{\}$ to denote anti-normal ordering, then
\begin{align}
\left\langle \left\{ \hat{x}^{2}\left(t\right)\right\} \right\rangle  & =\frac{1}{4}\left(1+e^{-2t}\right)\nonumber \\
\left\langle \left\{ y^{2}\left(t\right)\right\} \right\rangle  & =\frac{1}{4}\left(1+e^{2t}\right).
\end{align}
 In both cases there is a reduction in variance in the direction of
positive diffusion. If there is an initial vacuum state, then quadrature
squeezing occurs in $x$ in the forward time direction, with a variance
reduced below the vacuum level. Backward time squeezing occurs in
$y$, which has forward-time gain.

\subsection{Higher dimensional stochastic equation}

In the matrix notation used elsewhere, this means that we have $d=1/2$,
and
\begin{equation}
\bm{A}=\left[\begin{array}{c}
-x\\
y
\end{array}\right],
\end{equation}
with $\bm{c}=0$, so that the quantum dynamics occurs as the steady-state
of a higher dimensional equation:
\begin{equation}
\frac{\partial\bm{\phi}}{\partial\tau}=\ddot{\bm{\phi}}-\bm{\phi}+\bm{\zeta}\left(t,\tau\right),
\end{equation}
where $\left\langle \zeta^{\mu}\left(t,\tau\right)\zeta^{\nu}\left(t',\tau'\right)\right\rangle =\delta^{\mu\nu}\delta\left(\tau-\tau'\right)\delta\left(t-t'\right)$,
with boundary values such that:
\begin{align}
x\left(t_{0}\right) & =x_{0}\nonumber \\
y\left(t_{f}\right) & =y_{f}\nonumber \\
\dot{x}\left(t_{f}\right) & =-x\left(t_{f}\right)\nonumber \\
\dot{y}\left(t_{0}\right) & =y\left(t_{0}\right)\,.
\end{align}

These are called mixed boundary conditions. They are partly Dirichlet
(specified value), and partly Robin (specified linear combination
of value and derivative). Numerical solutions for the the squeezed
$x$ equations are given in Figs (\ref{fig:Quantum-squeeze_vs_tau_t})
and (\ref{fig:Quantum-squeeze_vs_t}), while those for the unsqueezed
$y$ equations are given in Figs (\ref{fig:Quantum-unsqueeze_vs_tau-t})
and (\ref{fig:Quantum-unsqueeze_vs_t}). The effects of sampling error
are seen through the two solid lines, giving one standard deviation
variations from the mean. Exact results are included via the dashed
lines.

\section{Summary\label{sec:Summary}}

The existence of a time-symmetric probabilistic action principle for
quantum fields describes a different approach to the understanding
of quantum dynamics. Neither imaginary time nor oscillatory path integrals
are employed. More generally, time evolution through a symmetric stochastic
action can be viewed as a dynamical principle in its own right. It
is equivalent to the traditional action principle of quantum field
theory. The advantage is that it is completely probabilistic, even
for real-time quantum dynamics. Although not all commonly used Hamiltonians
are included here, extensions to larger classes of quartic quantum
field Hamiltonians appear feasible.

A property of this method is that it can provide an ontological interpretation
of quantum mechanics. This quantum action principle can give a description
of a reality that underlies the Copenhagen interpretation. The picture
is of physical fields propagating both from the past to the future
and from the future to the past. This time-symmetric interpretation,
does not require a collapse of the wave-function. Such ontological
interpretations are different to hidden variable theories \citep{Bell1964},
which only allow causality from past to future. As a result, one can
have quantum features including vacuum fluctuations, sharp eigenvalues
and even Bell violations \citep{drummond2019q}, within a realistic
and local framework.

The present paper has focused on the conceptual basis of this approach,
and a proof of equivalence between quantum field dynamics and a time-symmetric
stochastic action. The nonlocal boundary conditions used are different
to local boundary conditions, and will not necessarily be equivalent
to every quantum state, which are defined locally in time. Conditional
boundaries are also possible in principle. These correspond to a larger
set of possible Q-functions and Hamiltonians, but are outside the
scope of the present paper.

The power of rapidly developing petascale and exascale computers appears
well-suited to these approaches. Enlarged spatial lattices and increased
parallelism are certainly needed. Yet this may not be as problematic
to handle as either exponential complexity or the phase problems that
arise in other approaches. It is intriguing that the utility of an
extra dimension is widely recognized both in general relativity and
quantum field theory. One may speculate that extending this action
principle to curved space-time may yield novel quantum theories. This
could lead to new approaches to unification.
\begin{acknowledgments}
PDD acknowledges useful discussions with Margaret Reid, Laura Rosales-Zarate
and Ria Joseph. He thanks the hospitality of the Institute for Atomic
and Molecular Physics (ITAMP) at Harvard University, and the Weizmann
Institute of Science through a Weston Visiting Professorship. This
work was also performed in part at Aspen Center for Physics, supported
by National Science Foundation grant PHY-1607611, and was funded through
an Australian Research Council Discovery Project Grant DP190101480,
and a grant from NTT Research. 
\end{acknowledgments}

\section*{}

\bibliographystyle{apsrev4-1}
\bibliography{Qbridgefull,RMP_Draft_references}

\end{document}